# Freeze-Casting of Porous Ceramics: A Review of Current Achievements and Issues

Sylvain Deville

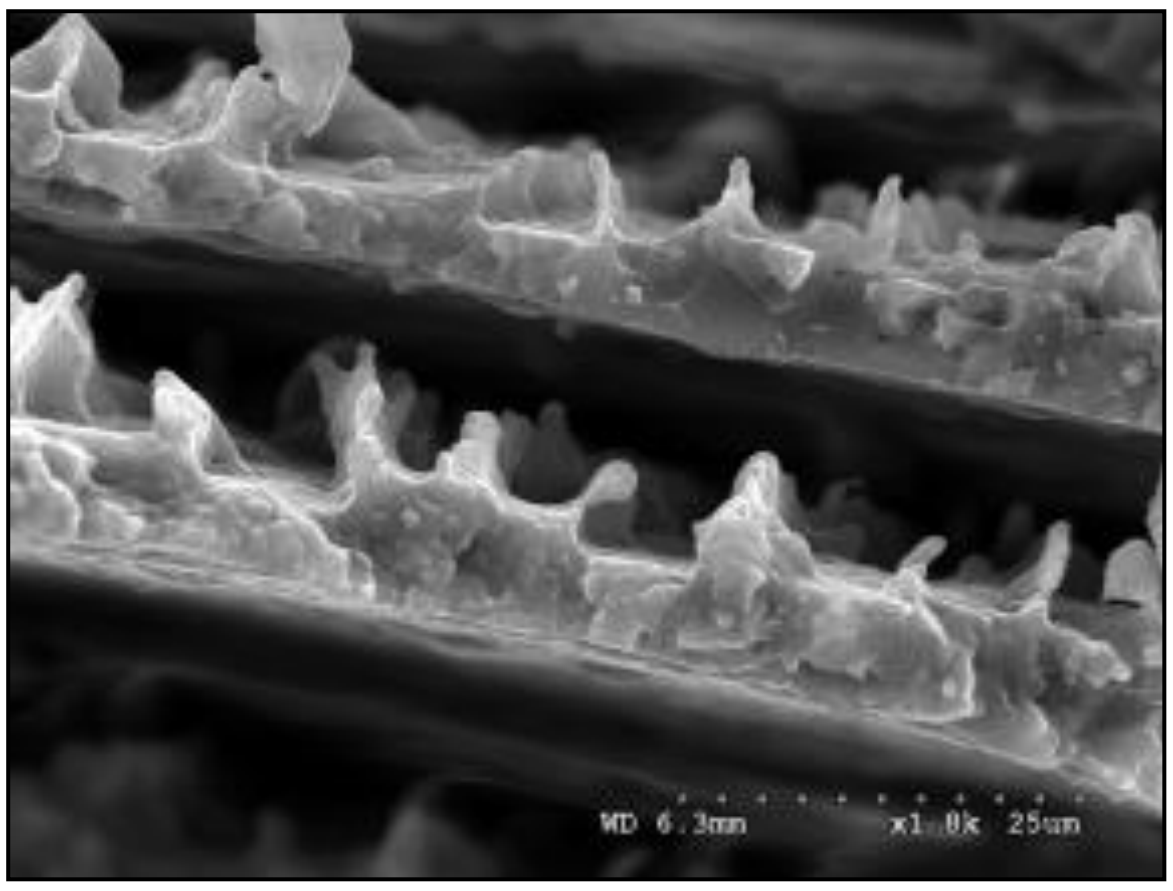

Freeze-casting of porous ceramics have seen a great deal of efforts during the last few years. The objective of this review is to provide a first understanding of the process as of today. This analysis highlights the current limits of both the understanding and the control of the process. A few perspectives are given, with regards of the current achievements, interests and identified issues.

# Freeze-Casting of Porous Ceramics: A Review of Current Achievements and Issues

Sylvain Deville

Laboratory of Synthesis and Functionnalisation of Ceramics

FRE2770 CNRS/Saint-Gobain CREE, 550, Avenue Alphonse Jauffret, BP 224, 84306 Cavaillon Cedex, FRANCE

Email : sylvain.deville@saint-gobain.com

## Abstract

Freeze-casting, the templating of porous structure by the solidification of a solvent, have seen a great deal of efforts during the last few years. Of particular interest are the unique structure and properties exhibited by porous freeze-casted ceramics, which opened new opportunities in the field of cellular ceramics. The objective of this review is to provide a first understanding of the process as of today, with particular attention being paid on the underlying principles of the structure formation mechanisms and the influence of processing parameters on the structure. This analysis highlights the current limits of both the understanding and the control of the process. A few perspectives are given, with regards of the current achievements, interests and identified issues.

## 1. Introduction

Although porosity in technical ceramics has been considered as problematic for a long time, the potentialities offered by porous ceramics are drawing considerably more attention today than just a few years ago. Cellular ceramics can be engineered to combine several advantages inherent from their architecture [1]: they are lightweight, can have open or closed porosity making them useful as insulators or filters, can withstand high temperatures and exhibit high specific strength, in particular in compression [2]. Typical processing methods include foam or wood replication [3-6], direct foaming [7] or extrusion. The full potential of cellular ceramics will only be achieved once a proper control of the size, shape and amount of porosity will be available. Although the control over the structure and functional properties of cellular ceramics is continuously improving, all processing routes suffer from an inherent limitation: every processing route is intrinsically limited to a narrow range of pores characteristics. In addition, removal of the pore forming agent can be a considerable problem, and efforts have been in put in developing processing routes with environmental friendly pore forming agents, yielding techniques such as gel casting [8, 9], direct foaming [7] or recent developments with particles-stabilized wet foams [10]. In the pursuit of such processing routes, freeze-casting has attracted considerably more focus in the last few years (Fig. 1). The technique consists of freezing a liquid suspension (aqueous or not), followed by sublimation of the solidified phase from the solid to the gas state under reduced pressure, and subsequent sintering to consolidate and densify the walls. A porous structure is obtained, with unidirectional channels in the case of unidirectional freezing, where pores are a replica of the solvent crystals. The technique seems to be rather versatile and the use of a liquid solvent (water most of the time) as a pore forming agent is a strong asset. Freeze-casting has also been developed as a near net shape forming route, yielding dense ceramics [11, 12]. This approach will not be discussed here, as our primary goal is to discuss the potentialities offered for porous ceramics.

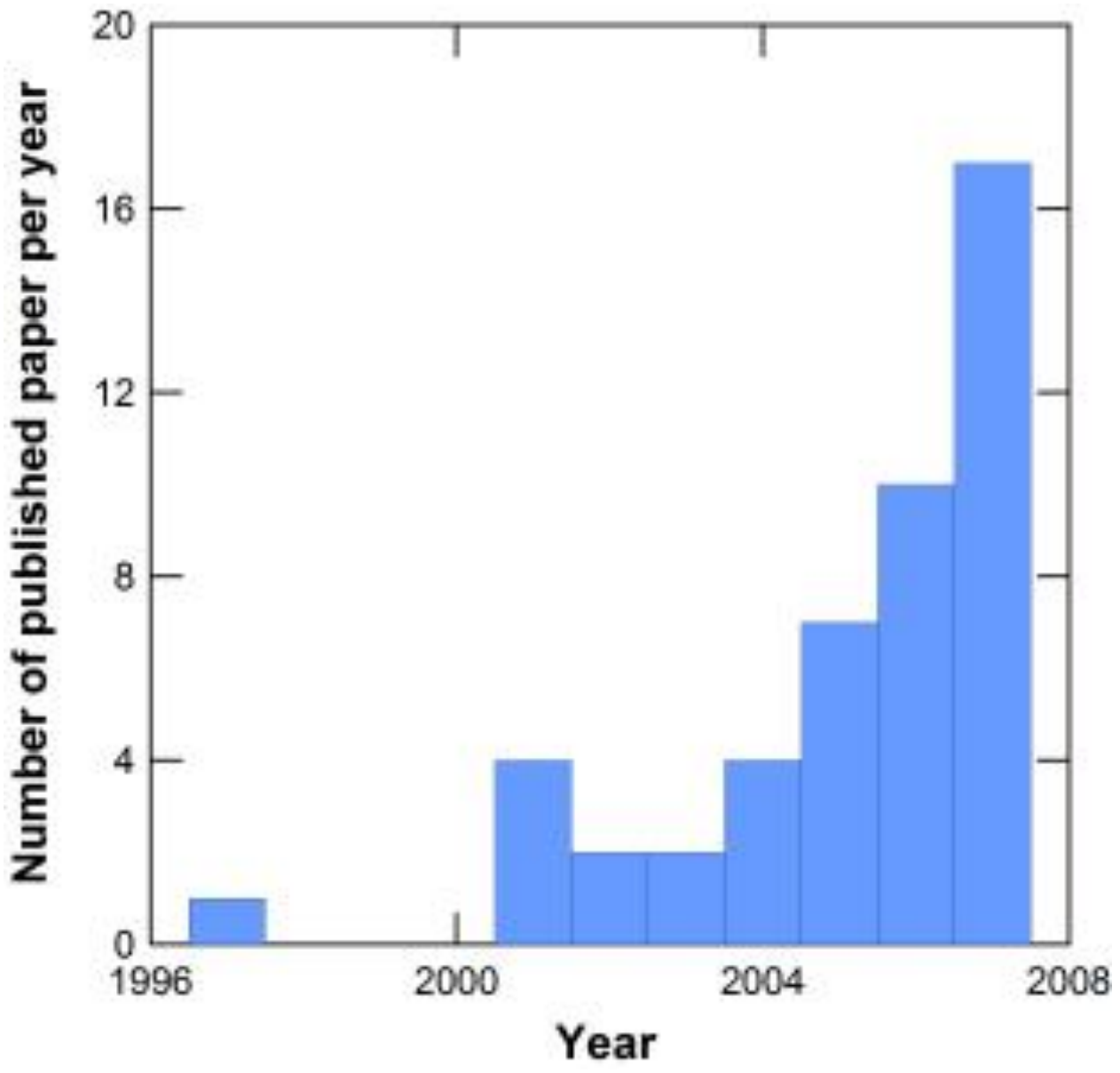

Figure 1: Evolution of papers published on freeze-casting of porous ceramics per year. For 2007, only papers published until November were taken into account.

The objective of this paper is to provide a first review of the results obtained up to date and to offer insights on the potentialities and limits of the technique. The current understanding and control over the processing route and final structure are described and discussed. The review is organized as follow. In a first part, the processing principles of freeze-casting are briefly described, before the materials processed to date are summarized. We then describe in details the structure and porosity characteristics of such materials, and provide the current understanding of the mechanisms controlling the formation of the porous structure. This understanding also defines the actual limitations of the technique, and provides perspectives for the future of freeze-casting, in terms of refinement of the process, control over the structure and tailoring of the functional properties.

## 2. Processing principles

Freeze-casting has first been developed as a near net shape forming technique [11, 12], yielding dense ceramics parts with fine replicate of the mould details. Any ice crystal being converted into porosity later on in the process, introducing large size defects largely unwelcome in ceramic applications, a great deal of efforts has been put in controlling and avoiding the formation of ice crystals. Only later on was it realized that the formation and growth of ice crystals could be a substantial benefit if properly controlled, yielding porous ceramics with a very specific porosity. The early work of Fukasawa on alumina [13] revealed the potentialities offered for porous ceramics, and a great deal of efforts has been subsequently put.

The technique consists of freezing a liquid suspension (aqueous or not), followed by sublimation of the solidified phase from the solid to the gas state under reduced pressure, and subsequent sintering to consolidate and densify the walls, leading to a porous structure with unidirectional channels in the case of unidirectional freezing, where pores are a replica of the solvent crystals.

In freeze-casting, the particles in suspension in the slurry are rejected from the moving solidification front and piled up between the growing cellular solvent crystals, in a similar way to salt and biological organisms entrapped in brine channels in sea ice [14]. The variety of materials processed by freeze-casting suggests that the underlying principles of the technique are not strongly dependent on the materials but rely more on physical rather than chemical interactions. The phenomenon is very similar to that of unidirectional solidification of cast materials and binary alloys, in particular when powders with small (submicronic) particles size are used, with the solvent playing the role of a fugitive second phase.

The processing can be divided in four steps (Fig. 2), and the corresponding experimental conditions will strongly depend on the chosen solvent.

1. **Preparation of the slurry**. This step is very similar to the preparation of slurries for conventional processing routes such as slip casting. The ceramic powder must be correctly dispersed in the liquid medium (the solvent), so that dispersant and plasticizer are often used. The temperature of the slurry must fall in the range were the solvent is liquid, room temperature in the case of water, but different temperature (60°C and 8°C) is necessary for respectively camphene-based and tert-butyl alcohol slurries. Moderate solid loading is used (10-40 vol.%), depending of the desired amount of total porosity. The stability of the suspension must be carefully controlled to avoid any segregation phenomenon taking place in the second stage, yielding gradients of density and porosity in the final materials. This can be particularly problematic for low solid loading. Finally, the presence of a binder is necessary, to provide green strength after sublimation. Though the solvent is playing the role of the structuring agent, binder and pore forming agent, it is nevertheless removed during the sublimation stage, so that green bodies collapse in absence of an organic binder. The role of additional additives will be discussed in the section devoted to the control of the structure.

2. **Controlled solidification of the slurry**. This is the critical stage where the structure is formed and the characteristics of the future porosity are determined. During this stage, continuous crystals of solvent are formed, under certain conditions, and grow into the slurry. Ceramic particles in suspension in the slurry are rejected by the moving solidification front, concentrated and entrapped in-between the crystals. To induce this natural segregation phenomenon, the slurry is poured in a mould, which undergoes isotropic or anisotropic cooling to induce homogeneous or directional solidification. Several devices, also used to process porous polymers by freeze-casting, have been designed to provide a more or less elaborated control of the solidification conditions [13, 15-20]. The solidification conditions are dictated by the initial choice of the solvent. Low temperatures (<0°C) are required when using water, while room temperature are sufficient when using camphene, its solidification point being around 44-48°C. The device should also accommodate the solidification shrinkage; negative (shrinkage) in the case of camphene (-3.1%) and positive (expansion) in the case of water (+9%). The cooling conditions will largely dictate the characteristics of the growing solvent crystals and hence the final characteristics of the porosity.
3. **Sublimation of the solvent**. Once complete solidification of the sample is achieved, the sample is kept at conditions of low temperature and reduced pressure, conditions dictated by the physical properties of the solvent. Under these sublimation conditions, the solidified solvent is converted into the gas state. Porosity is created where the solvent crystals were, so that a green porous structure is obtained; the porosity is a direct replica of the solidified solvent structure. When using water, a conventional freeze-dryer can be used. In the case of camphene, the vapor pressure of 1.3 kPa (just below the melting temperature) is high enough to allow sublimation at room temperature, so that no specific equipment is required.

4. **Sintering or densification of the green body**. Once the solvent has been totally removed, the obtained green body can be sintered with conventional sintering technique. The low strength of the green body prevents any use of pressure assisted sintering. The low amount of organic binder (usually <5%) does not require the presence of a special and often problematic binder burnout process. During the sintering stage, microporosity can be removed from the ceramic walls, but the macroporosity created by the solvent crystals is retained.

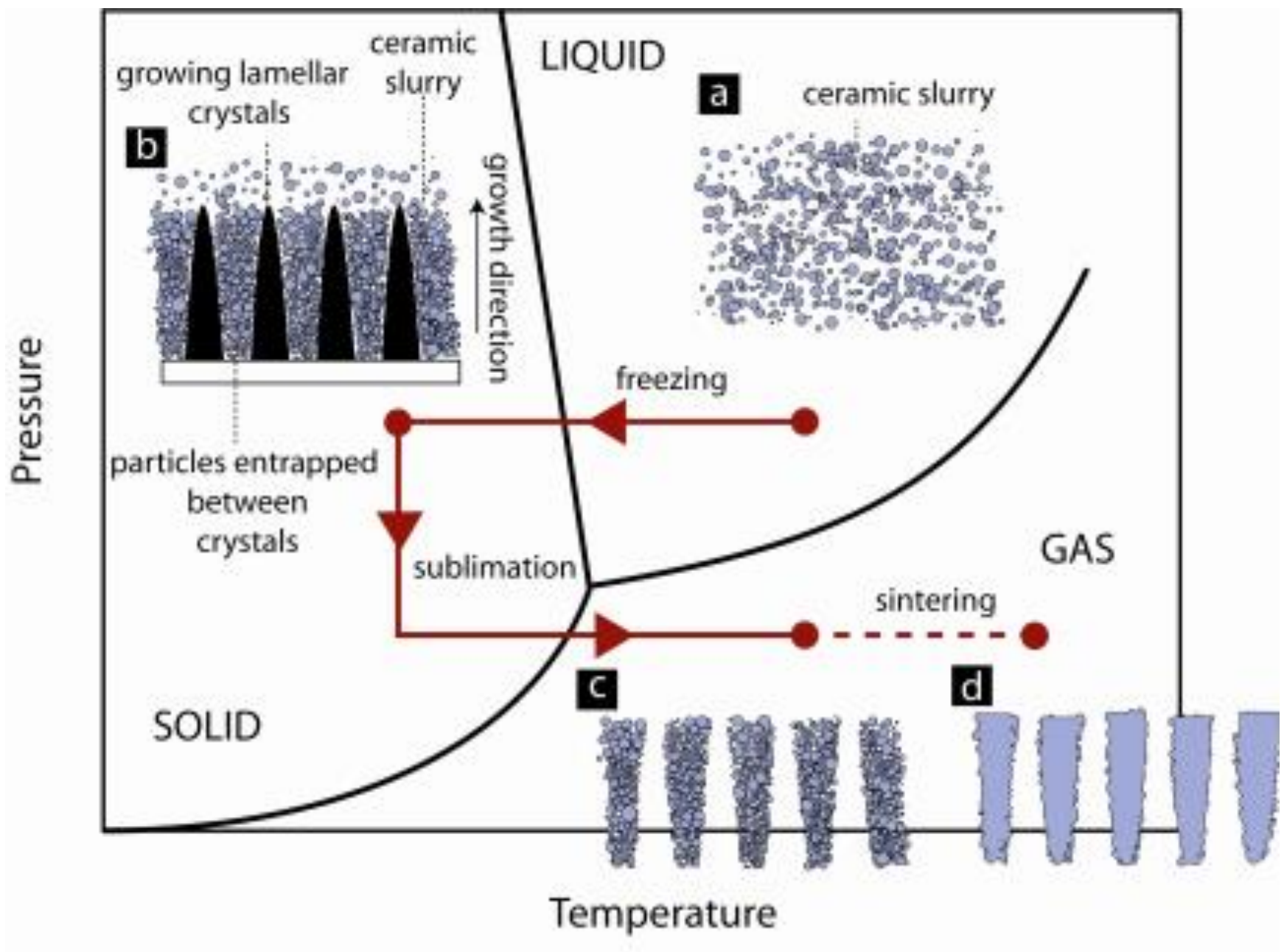

Figure 2: The four processing steps of freeze-casting: slurry preparation, solidification, sublimation and sintering.

## 3. Materials

A wide variety of ceramic materials have already been tested, including alumina [13, 17, 21-28], hydroxyapatite and tricalcium phosphate [29-36], NiO-YSZ [20], Ni-YSZ [37], yttria-stabilised zirconia [16, 38], titanium dioxide [39], silicon nitride [40, 41], PZT-PZN [42], mullite[34] (from alumina gel with ultrafine silica) [43], glass [44, 45], silica [18, 46, 47], silica-alumina [48], clay [49], LSCF-CGO [19], MgO (from magnesium sulfate) and silicon carbide [50]. Porous polymer-ceramic composites have also been processed [51] using the technique. The variety of materials processed by the technique suggests that the underlying principles dictating the structure formation mechanisms rely on physical interactions, making the process a versatile one.

An interesting variation was reported by Yoon *et al.* [52]. Highly porous silicon carbide with unidirectional porosity was processed at low temperature, starting from freeze-casted green bodies of silicon carbide precursor (or a mix of powder and precursor[53]), polycarbosilane in that case, and using camphene as a solvent. The green bodies were pyrolyzed at 1400°C under argon, a consolidation temperature much lower than that used for the recrystallization of silicon carbide (typically 2200°C).

| Study | Materials | Powder particle size | Solvent | Cooling Setup | Cooling conditions | Cooling Directionality | Pore size range* (short axis) | Sintering temperature | Comments | Reference |
|---|---|---|---|---|---|---|---|---|---|---|
| Fukasawa *et al*, 2001. | Alumina | | Water | Mould with bottom part immersed in freezing bath | -20°C to -80°C | Yes | 40 μm | 1400-1600°C | | [21] |
| Fukasawa *et al*, 2001. | Alumina | | Water | Mould with bottom part immersed in freezing bath | -50°C | Yes | 30 μm | 1400-1550°C | | [13] |
| Donchev *et al.*, 2005 | Alumina | 700 nm | Silica sol | Cold plate | -30°C | Yes | 80 μm | No sintering | | [22] |
| Koch *et al.*, 2003 | Alumina | 700 nm | Silica sol | Metallic mould cooled from one side | Ramp down to -40°C | Yes | 50 μm | 1100-1500°C | | [23] |
| Araki *et al.*, 2005 | Alumina | 400 nm | Camphene | Mould with bottom part immersed in freezing bath | Cooling from 55°C down to RT | Yes | 5-20 μm | 1600°C | | [24] |
| Deville *et al.*, 2006, 2007 | Alumina | 400 nm | Water | Temperature gradient | Cooling from RT to -100°C | Yes | 2-200 μm | 1500°C | | [17, 25] |
| Koh *et al.*, 2006 | Alumina | 300 nm | Camphene | Cold water bath, 20°C | Cooling from 60°C | No | 5-40 μm | 1400°C | Polystyrene addition | [26] |
| Nakata *et al.*, 2005 | Alumina | 300 nm | Water | Isotropic and anisotropic cooling, fixed temperature | -20°C | Yes/No | 2-200 μm | 1500°C | | [27] |
| Araki *et al.*, 2004 | Alumina | 400 nm | Camphene-naphthalene | Mould at RT | Cooling from 60°C | No | 0-10 μm | 1600°C | | [28] |
| Chen *et al.*, 2007 | Alumina | 200 nm | Tert-butyl alcohol | Mould with bottom part immersed in freezing bath | 0°C and -15°C | Yes | 50 μm | 1500°C | Additional gelation step | [54] |
| Shanti *et al.*, 2006 | Alumina | 400 nm | Camphene | Mould at RT | RT | Yes | 3-20 μm | 1600°C | | [55] |
| Fukasawa *et al.*, 2002 | $Si_3N_4$ | 550 nm | Water | Mould with bottom part immersed in freezing bath | -25°C and -80°C | Yes | 50 μm | 1700-1850°C | Filtering properties tested | [40] |
| Fukasawa *et al.*, 2002 | $Si_3N_4$ | 550 nm | Water | Mould with bottom part immersed in freezing bath | -50°C and -80°C | Yes | 30 μm | 1700-1850°C | | [41] |
| Ding *et al.*, in press | Mullite | | Alumina and silica sol | Cooling from RT? | -20°C | ? | 100 μm | 1400-1600°C | | [43] |
| Lee *et al.*, in press | PZT-PZN | | Camphene | Mould at RT | Cooling from 60°C | No | 10-50 μm | 1200°C | Piezoelectric properties tested | [42] |
| Koh *et al.*, 2007 | Ni-YSZ | 300 nm | Camphene | Mould at fixed temperature | 20°C | No | 30 μm | 1100-1400°C | | [37] |
| Moon *et al.*, 2003 | NiO-YSZ | | Water | Mould at fixed temperature | -30°C | Yes | 10 µm | 1400°C | Radial porosity | [20] |
| Moon *et al.*, 2002 | LSCF-CGO | | Water | Mould with bottom part immersed in freezing bath | -80°C | Yes | 10 µm | 1350°C | Thin film | [19] |
| Tang *et al.*, 2005 | SiC | 16 µm | Water | Mould with bottom part immersed in freezing bath | -40°C | Yes | 300 µm? | No sintering | | [50] |
| Yoon *et al.*, 2007 | SiC | Precursor, precursor+ powder (300 nm) | Camphene | Mould at fixed temperature | 20 to -40°C | Yes | 2-30 µm | 1400°C | Starting from polycarbosilanes | [52, 53] |
| Hwang *et al.*, 2006 | Clay | | Water | Mould with bottom part immersed in freezing bath | -70°C | Yes | 20 µm? | 800°C | Extremely high porosity (>94%) | [49] |
| Kisa *et al.*, 2003 | Silica | 20 nm | Silica sol | Single crystal wafer at fixed temperature | -196°C | Yes | 3-10 µm | 600-1000°C | | [46] |
| Mukai *et al.*, 2004 | Silica | | Silica gel | Dipping of mould in liquid $N_2$ | -196°C | Yes | 10 µm | No sintering necessary | | [18] |
| Sofie *et al.*, in press | 8YSZ | 300 nm | Camphene | Cold support, -25°C | Cooling from 55°C | Yes | 2-25 µm | 1400°C | Freeze-tape casting with porosity gradient | [16] |
| Ren *et al.*, in press | $TiO_2$ | 150 nm | Water | Cold support, -18°C | -18°C | Yes | 15 µm | 1000°C | Freeze-tape casting with porosity gradient | [39] |
| Bettge *et al.*, 2005 | YSZ | | Water | Mould with bottom part immersed in freezing bath | -40°C | Yes | 15-30 µm | ? | | [38] |
| Koh *et al.*, 2006 | 8YSZ | | Camphene | Mould at RT | RT | Yes | Gradient | 1400°C | Dense/porous bilayer | [56] |
| Nishihara *et al.*, 2005 | Silica | | Silica sol | Dipping in cold bath at fixed temperature | -196°C and -60°C | Yes | 3-40 µm | 605 and 905°C | | [47] |
| Nishihara *et al.*, 2006 | Silica-alumina | | Silica sol | Dipping in cold bath at fixed temperature | -196°C | Yes | 20 µm | 550°C | | [48] |
| Deng *et al.*, in press | Glass | 6.5 µm | Water | Mould at fixed temperature | -55°C | Yes | 100-150 µm | 620°C | Nano $TiO_2$ thin film coating | [44] |
| Song *et al.*, 2006 | Bioglass | | Camphene | Mould at 20°C | Cooling from 60°C | No | 20-40 µm | 700-1100°C | | [45] |
| Deville *et al.*, 2006 | HAP | 2 µm | Water | Temperature gradient | Cooling from RT | Yes | 15-40 µm | 1250-1350°C | | [30, 34, 35] |
| Lee *et al.*, 2007 | HAP | 10 µm? | Camphene | Mould at fixed temperature, 20°C | Cooling from 60°C | No | 20-40 µm | 1250°C | | [31] |

| | | | | | | | | | |
|---|---|---|---|---|---|---|---|---|---|
| Yoon *et al.*, 2007 | HAP | 10 μm? | Camphene | Mould at fixed temperature, 0 to 35°C | Cooling from 60°C | No | 80-220 μm | 1250°C | [29, 32] |
| Moritz *et al.*, in press | HAP | 2 μm | Water | Dipping in cold bath | -19°C | No | 20-100 μm | 1350°C | [33] |
| Suestsaga *et al.*, 2007 | HAP | 100 nm | Water | Mould with bottom part immersed in freezing bath | -196°C | Yes | 50 μm | 1200°C | [36] |

Table 1: Summary of materials, processing conditions and porosity. * in the case of water-based slurries, the porosity in lamellar, and the pore dimensions in plane can be defined by a long and a short axis. The highest number indicated here corresponds to the dimension of the short axis. See reference [30] for details.

## 4. Structure and Properties

### *4.1 Structure*

**Macroporosity**

The porosity of the sintered materials is a replica of the original solvent crystals. A variety of pores morphology can be obtained, depending on the choice of the solvent, slurries formulation and the solidification conditions (Fig. 3). Since the solidification is often directional, the porous channels run from the bottom to the top of the samples. Homogeneous freezing (i.e., cooling of the fingers at constant rate starting from room temperature) results in a more homogeneous ice nucleation [15] leading to a lamellar porous architecture (Fig. 3b), with long range order, both in the parallel and perpendicular directions of the ice front. After sintering, the ceramics walls can be completely dense with no residual porosity, depending on the sintering conditions.

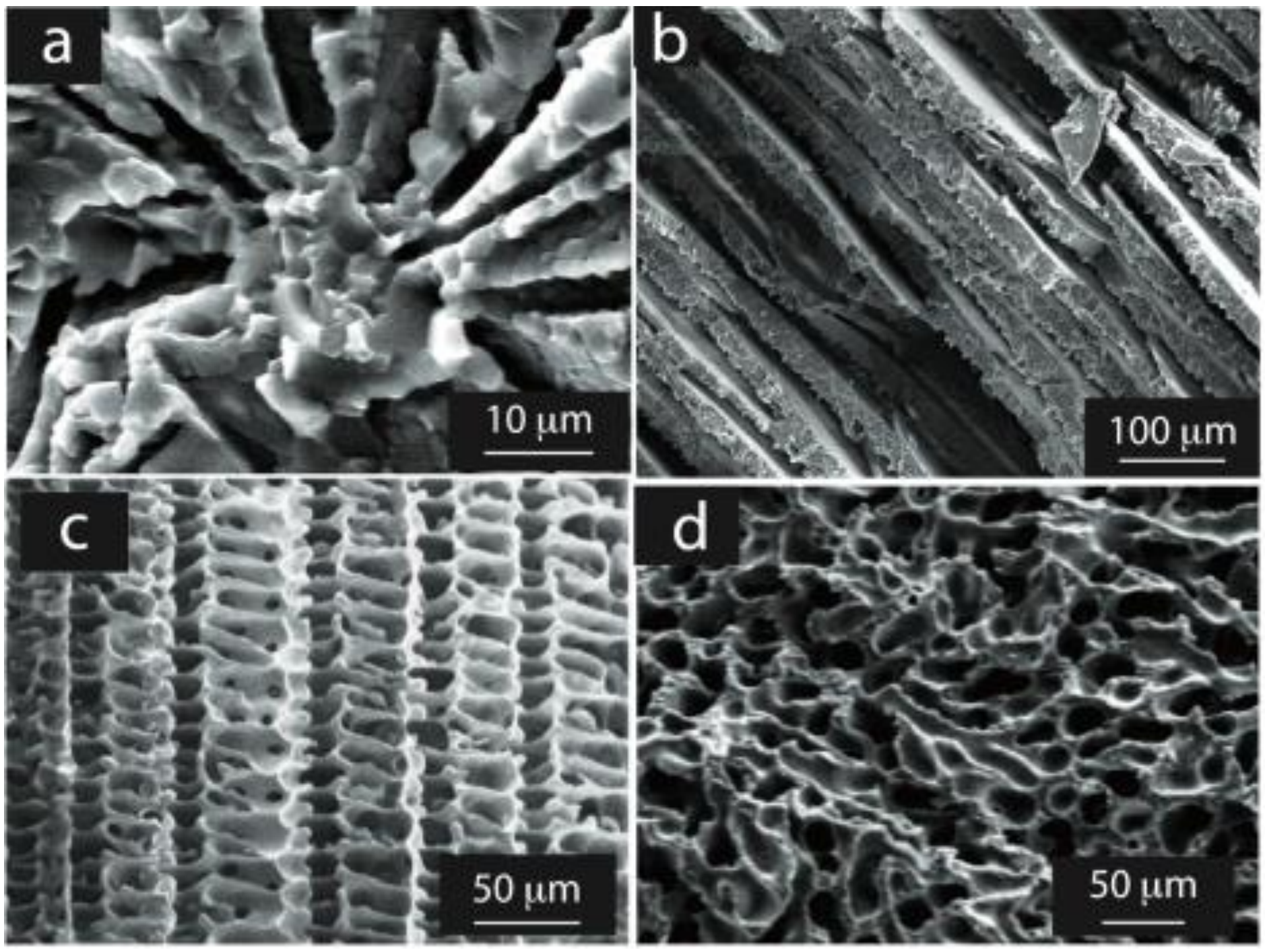

Figure 3: Typical microstructures obtained by freeze-casting (a) porous alumina using an hypoeutectic camphor/naphthalene as a solvent [28] (after K. Araki, Room-Temperature Freer-Casting for Ceramics with Nonaqueous Sublimable Solvent Vehicles, Journal of the American Ceramic Society, Blackwell Publishing, with permission) (b) porous alumina using water as a solvent [25] (c) porous silicon carbide using polycarbosilane as a precursor and camphene as a solvent [52] (after B. H. Yoon, Highly Aligned Porous Silicon Carbide Ceramics by Freezing Polycarbosilane/Camphene Solution, Journal of the American Ceramic Society, Blackwell Publishing, with permission) and (d)

porous alumina using camphene as a solvent [24] (after K. Araki, Porous Ceramic Bodies with Interconnected Pore Channels by a Novel Freeze-Casting Technique, Journal of the American Ceramic Society, Blackwell Publishing, with permission).

In the particular case of water being used as a solvent, the microstructure is lamellar, with lamellar channels between the ceramics walls. This particular morphology can be understood with reference to the basic crystallographic (Fig. 4a) and crystal growth characteristics of ice. The ice front velocity parallel to the crystallographic c axis is $10^2$ to $10^3$ times lower than perpendicular to this axis (Fig. 4b). After the transition to columnar ice occurred, ice platelets with a very large anisotropy can then be formed very fast with ice growing along the a-axes, while the thickness (along the c-axis) remains small. The freezing process is easier for crystals whose c-axes are perpendicular to the temperature gradient, such that growth along the gradient can occur in the a- or b-direction. The crystals with horizontal c-axes will therefore grow at the expense of the others and continue to grow upward, in an architecture composed of long vertical lamellar crystals with horizontal c-axes. In the final structures, the direction perpendicular to the lamellae corresponds thus to the original c-axis of ice crystals (Fig. 4c).

Similar explanations can be invoked to explain the morphology of the porosity in the sintered ceramics. In the case of camphene, the solidification of liquid camphene leads to the formation of clearly defined dendrites (Fig. 5), which are reflected in the final structures. Prismatic channels were obtained using tert-butyl alcohol as a solvent [54]. Using other solvents will likely provide different types of morphologies.

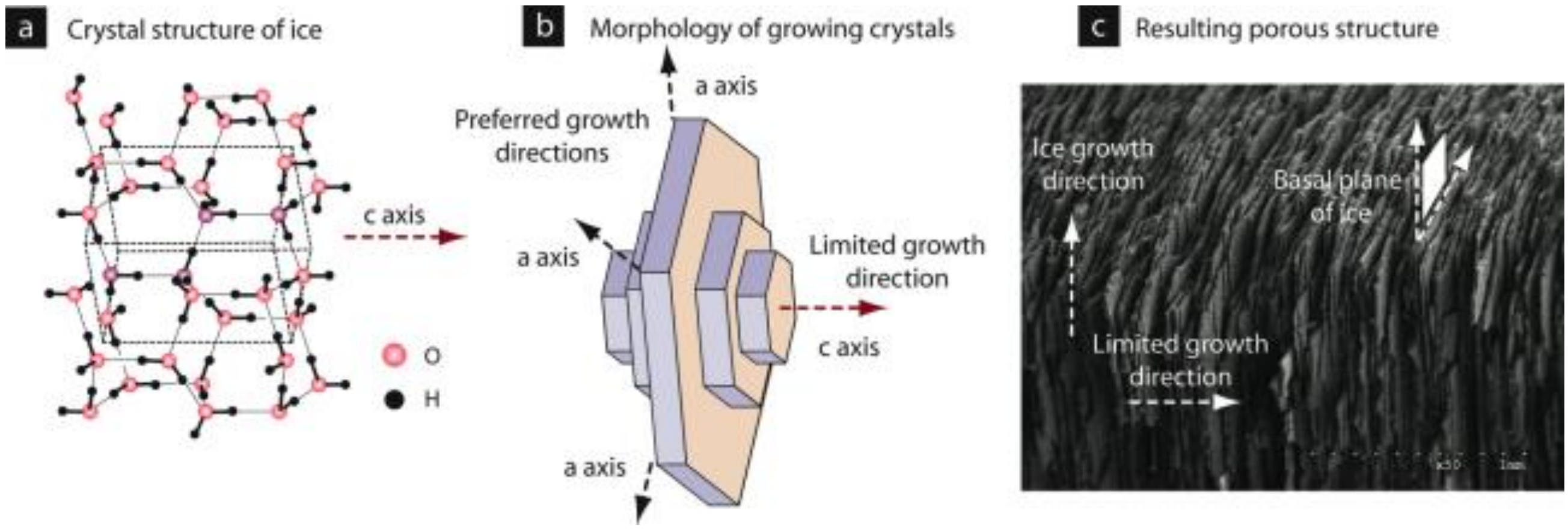

Figure 4: Crystal structure of ice (a) and anisotropy of crystal growth kinetics (b), leading to lamellar ice crystals. The anisotropy of the growth kinetics is reflected in the final porous structures (c) obtained after sublimation and sintering. The direction perpendicular to the ceramic platelets corresponds to the limited growth direction of ice crystals.

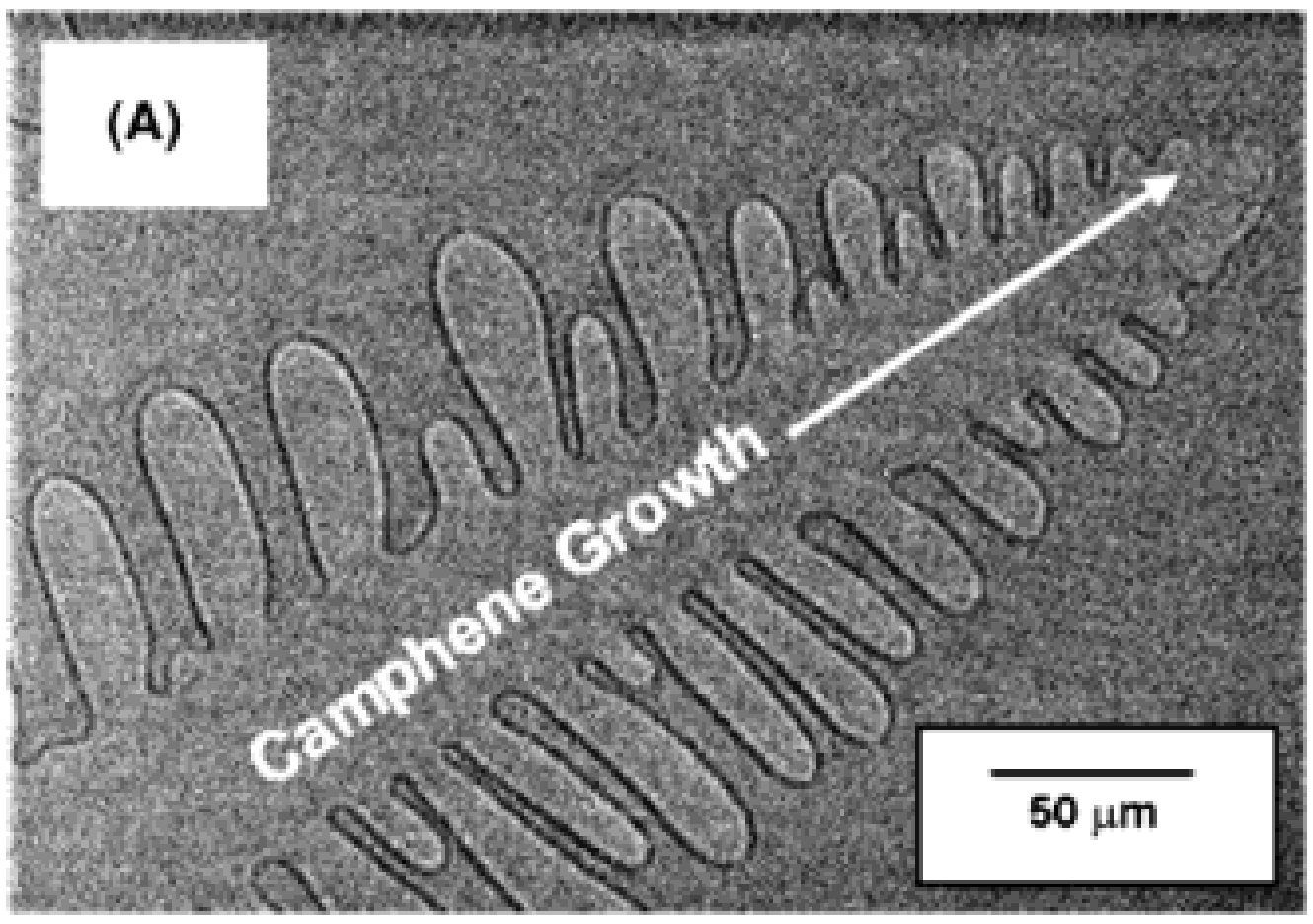

Figure 5: Solidified camphene dendrites [52], leading to the formation of a dendritic porous structure. After B. H. Yoon, Highly Aligned Porous Silicon Carbide Ceramics by Freezing Polycarbosilane/Camphene Solution, Journal of the American Ceramic Society, Blackwell Publishing, with permission.

**Orientation of macroporosity**

The pore channels can be oriented, depending on the solidification conditions. In most of the cases, the mould in which the slurry is initially poured is left with its bottom in part in contact with a cold surface. The solvent crystals are therefore solicited to grow vertically, along the direction of the imposed thermal gradient. However, different thermal gradients can be imposed, to induce a different anisotropy in the structure. A neat example of such control can be found in the study of Moon *et al.* [20]. Ice was stimulated to grow in the radial direction of the setup, from the inner surface of the metal cylinder to its centre region where a Teflon rod was placed. The final tubular structure exhibits a radially oriented porosity (Fig. 6), extending from the inside towards the outside of the tube. This type of structure could ideally be used for SOFC.

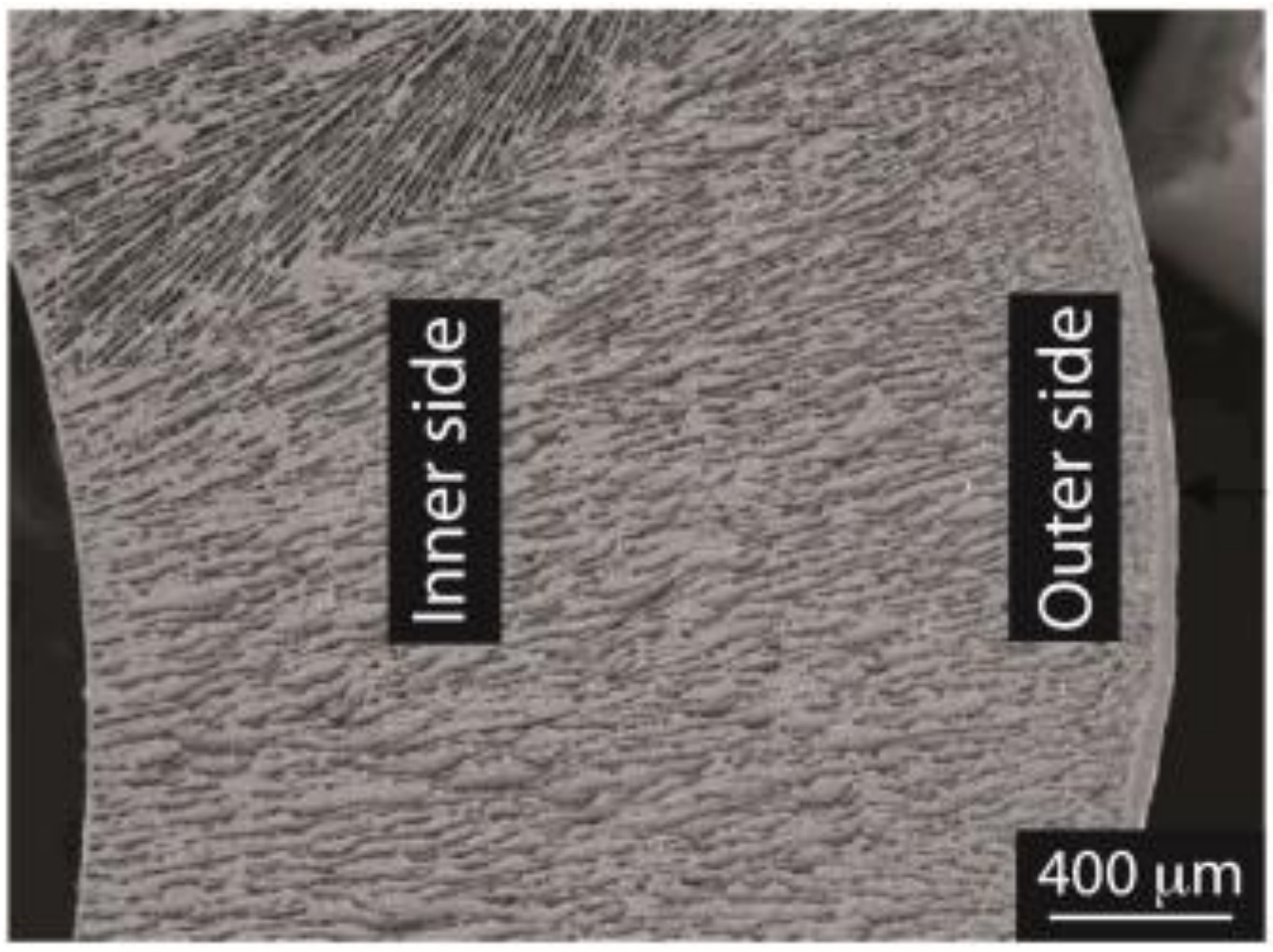

.

Figure 6: Radial porosity in a tubular structure [20]. Reprinted from Materials Letters, 7, Ji-Woong Moon, Hae-Jin Hwang, Masanobu Awano and Kunihiro Maeda, Preparation of NiO–YSZ tubular support with radially aligned pore channels, 1428-1434, Copyright (2003), with permission from Elsevier.

**Surface roughness of walls**

The surface of the lamellae exhibits a particular topography, with dendritic-like features, running in the solidification direction (Fig. 7a-b). These features are homogeneous in size and distribution, but their relative size varies with the freezing conditions, the nature of the solvent, the characteristics of the starting powders and the sintering conditions (atmosphere). In the particular case of silicon nitride (Fig. 7c), the microstructure reveals the presence of elongated fibrous grains protruding from the walls. This particular feature is not related to the freezing conditions, but only appears after sintering at rather high temperatures, and is believed to result from a vapor-solid transformation taking place during sintering [41]. A different roughness is obtained when camphene is used instead of water (Fig. 7d). In general, since the roughness is directly related to the morphology of the solvent crystal, every solvent will yield a different roughness.

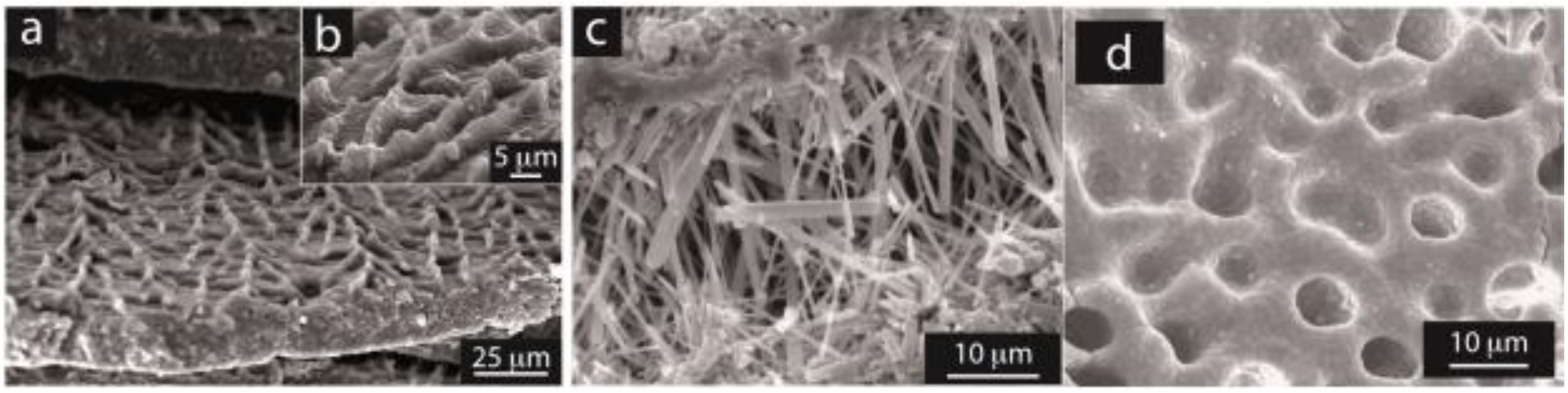

Figure 7: Dendritic surface of (a) alumina using water as a solvent [25] (b) silicon nitride using water as a solvent [41] (after T. Fukasawa, Synthesis Of Porous Silicon Nitride With Unidirectionally Aligned Channels Using Freeze-Drying Process, Journal of the American Ceramic Society, Blackwell Publishing, with permission) and hydroxyapatite using camphene as a solvent [32] (after B. H. Yoon, Generation of Large Pore Channels for Bone Tissue Engineering Using Camphene-Based Freeze Casting, Journal of the American Ceramic Society, Blackwell Publishing, with permission).

**Ceramic bridges**

Other features of these structures are the trans-lamellar ceramic bridges, of two types. The first type just corresponds to overgrown dendrites that eventually bridge the gap between two adjacent lamellae (Fig. 8). The second type is found in samples made from concentrated slurries. These numerous fine features with often-tortuous morphologies are locally bridging the gap between two adjacent lamellae. The morphology of these features is sometime quite different to that of the dendrites covering the ceramics lamellae, suggesting another formation mechanism. It has been proposed that they might be formed because of the specific conditions encountered during the slow freezing of highly concentrated solutions. The interaction of inert particles and a moving solidification front has been investigated for suspensions with low particles content. In such a case, the interaction between particles is not taken into account, which considerably simplifies the associated formalism. In the case of highly concentrated solutions, the particle-particle interactions cannot be ignored anymore. Eventually, it might considerably affect the pattern formation mechanisms. It has previously been shown that the particles themselves may induce morphological transitions, such as dendrites tip splitting or healing during growth before being captured [57]. Ceramic bridges between lamellae may arise from local ice crystal tip splitting and engulfment of particle agglomerates created by particles repelled from the ice-water interface and subsequent tip healing. Depending on the magnitude of tip splitting/healing, the entrapped ceramic particles might not bridge completely the gap. The phenomenon appears to be dependent on the nature of the solvent and will largely depend on the morphology of the growing dendrites.

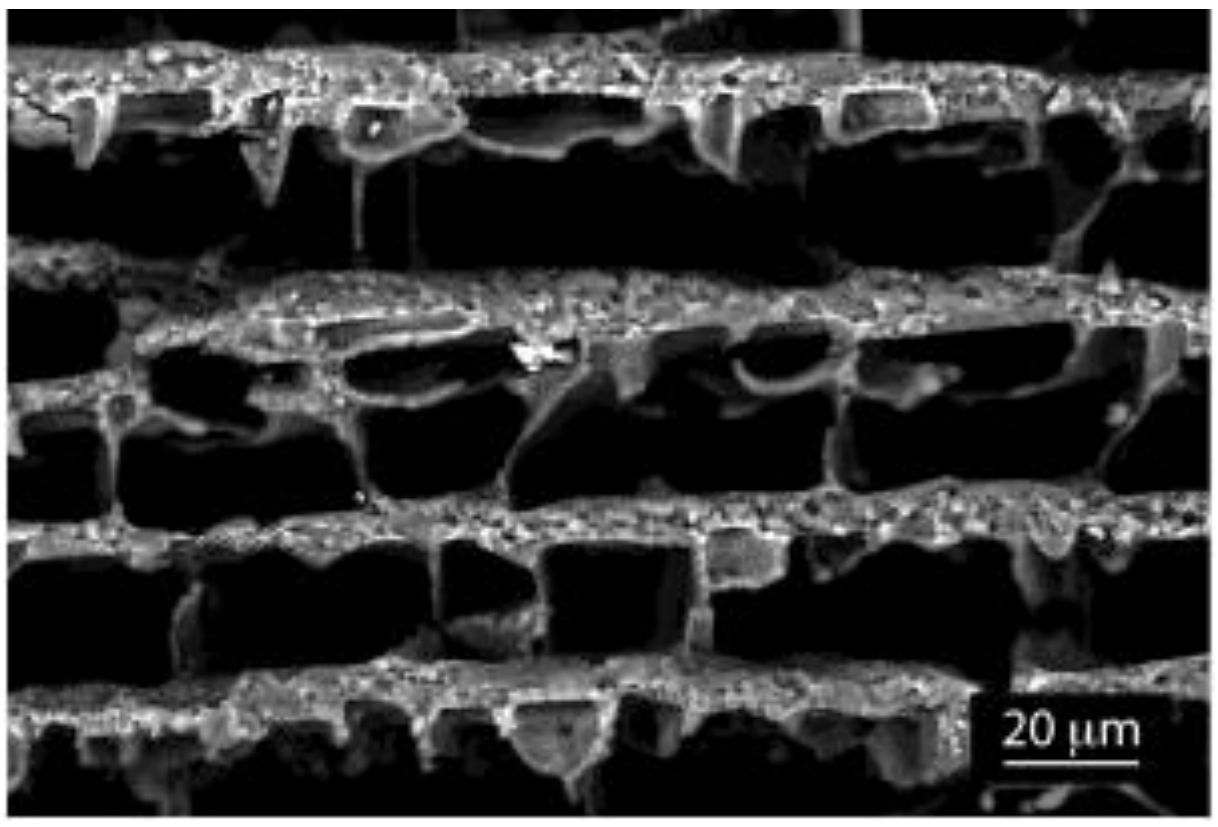

Figure 8: Ceramic bridges in porous lamellar alumina.

### *4.2 Properties*

Few properties of freeze-casted ceramics have been measured so far, as most of the attention was paid to the control of the structure and the processing conditions. Preliminary reports can nonetheless be found on one of the most critical property of cellular ceramics: the compressive strength. The strength is of particular importance in the case of intrinsically weak ceramics such as calcium phosphate, being considered as potential candidates for bone replacement applications. Dramatic improvements of compressive strength of hydroxyapatite (HAP) were reported [17], using water as a solvent (i.e., lamellar architecture). Although for high porosity content (typically >60 vol%) the strength (16 MPa) is comparable to that reported in the literature, it increases rapidly when the porosity decreases, reaching 65 MPa at 56% porosity and 145 MPa at 47% porosity. Values obtained for these samples are well above those reported so far in the literature. In fact, the strength of the porous lamellar HAP is close to that of compact bone. Such high values allow considering the potential of these materials for some load-bearing applications. Porous hydroxyapatite was also processed by freeze-casting [29, 31, 32] using camphene as a solvent, resulting in very different pores morphologies. Interestingly, the trend followed by the compressive strength (Fig. 9) seems to be highly dependent on the morphology of the pores. With camphene, the pores are dendritic and the structure somewhat close to usual cellular ceramics. With water, the pores are lamellar, and the compressive strength reaches much greater values, in particular for lower porosity values, with dramatic improvements observed. A possible explanation could be the strong anisotropy of the structure in the loading direction and the presence of inorganic bridges between the ceramic layers, which might prevent Euler buckling of the ceramic layers. Further investigations are nonetheless needed to rationalize these observations and confirm these trends.

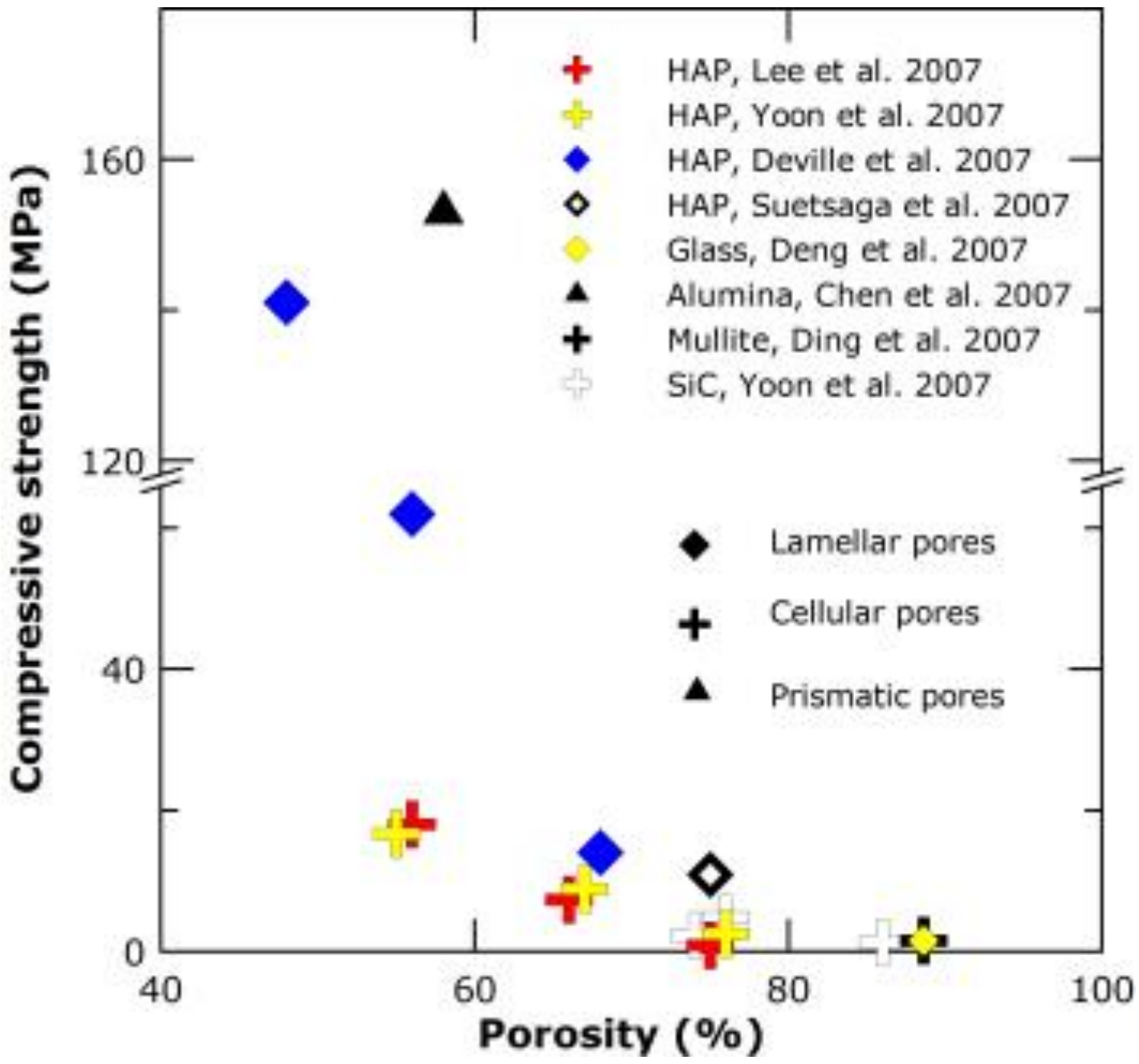

Figure 9: Compressive strength vs. total porosity, data from references [17, 31, 32, 36, 43, 44, 54]. The compressive strength seems to be highly dependent of the morphology of the pores, with much higher values measured for freeze-casted ceramics with lamellar and prismatic morphologies, as previously reported in reference [17].

## 5. Formation and control of the structure

Parameters affecting the final structure can be found in all the processing stages: slurry formulation and preparation (including the characteristics and properties of the starting powders), solidification and sintering. A few parameters are very specific and related to the core of the process: the freezing of the slurry. The solidification stage is the most critical with regards of the final porous structure; most of the features of the porosity will be created during this stage.

The formation of regular patterns is a common feature of many solidification processes, such as eutectic growth or unidirectional solidification of two-phases systems [17, 58]. Control of the regularity and size of the patterns is often a key issue with regards to the final properties of the materials. Many of the features of the freeze-casted porous ceramics can be understood by applying generic principles of solidification processes. In particular, the physics of water freezing has drawn the attention of scientists for a long time, for implications in fields as diverse as cryopreservation [59], cleaning of pollutants [60] or even polar ice formation [14, 61].

### *5.1 Formation of the structure: the interaction between the solidification front and the ceramic particles*

In order to obtain ceramic samples with a porous structure, two requirements must be satisfied:

(1) The ceramic particles in suspension in the slurry must be rejected from the advancing solidification front and entrapped between the growing ice crystals. This aspect can be understood considering the interaction between the solvent solidification front and the particles in suspensions. A simple thermodynamic criterion can be used in a first approach. The thermodynamic condition for a particle to be rejected by the solidification front is that there is an overall increase in surface energy if the particle is engulfed by the solid, i.e.,

$$\Delta\sigma = \sigma_{sp} - (\sigma_{lp} + \sigma_{sl}) > 0 \text{ (Eq. 1)}$$

where $\sigma_{sp}$, $\sigma_{lp}$ and $\sigma_{sl}$ are the interfacial free energies associated with the solid-particle, liquid-particle and solid-liquid interface respectively. When this criterion is satisfied and the particles rejected by the front, a liquid film should exist between the solidification front and the particle in order to maintain the transport of molecules towards the growing crystal. When the velocity of the front increases, the thickness of the film decreases. There is a critical velocity, $v_c$, for which this thickness is not enough to allow the necessary flow of molecules to keep the crystal growing behind the particle, that becomes then encapsulated by the solid. A large amount of theoretical and experimental work has been addressing this problem, and several expressions derived for the critical velocity. A few examples can be found in the references [62-64]. The main physical parameters to be taken into account includes the viscosity of the liquid, the particle size, the thickness of the film and the variation of free energy defined in equation 1. Although the complexity of the system leaves enough room for discussing the relevance of the various models, they are nonetheless useful to understand the influence of the physical parameters on the behavior of the systems and the morphology of the materials resulting from these interactions. These aspects are discussed in the next part.

(2) The ice front must have a non-planar morphology. Indeed, if the front is planar and the particles rejected, all the particles are collected on one side of the sample once solidification is achieved. This effect is being used in the purification of pollutants [60]. However, to form porous structures, particles redistribution must occur; the particles must be rejected from the solidification front and collected between the arms of the solidification front. The morphology of the front will then dictate the architecture of the final materials. At the very beginning of solidification (Fig. 10a), the interface (the front) is planar, and must somehow undergo a transition towards an irregular morphology, i.e., cellular, lamellar, or even more complex dendritic morphologies. This transition can be triggered by different mechanisms. One mechanism is the inherent thermodynamic instability of the interface (Fig. 10b), also known as a Mullins-Sekerka instability [65]. The development of the instability is based on supercooling effects, building up with solute rejection ahead of the interface. Another mechanism is related to the presence of the particles. In that case, the instability is due to the reversal of the thermal gradient in the liquid ahead of the interface and behind the particle (Fig. 10c) [66, 67]. It is not clear at this point which of these mechanisms is dominating. Further discussion can be found in the reference [25].

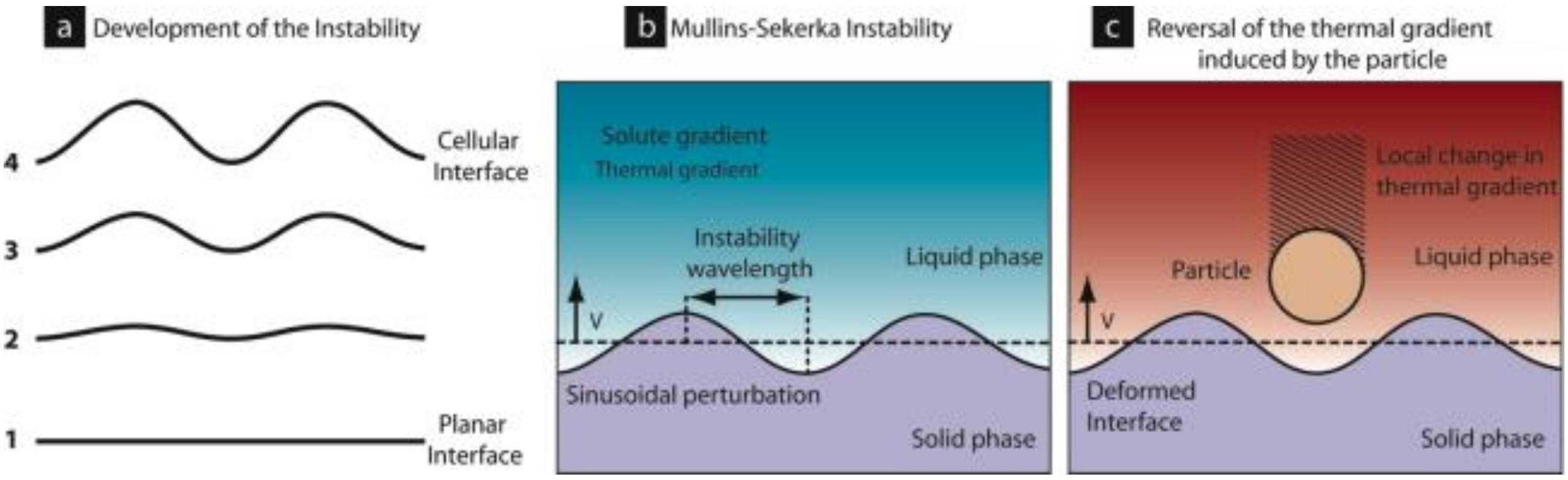

Figure 10: Destabilization of the interface and possible mechanisms triggering this morphological transition.

### 5.2 *Influence of processing and physical parameters*

**Solvent** – The choice of the solvent will be crucial both in regards of the processing conditions and on the characteristics of the structure that is desired. If freeze-casting was originally developed using water as a solvent, alternative solutions are considered today. Requirements for such alternative includes solidification temperature, viscosity of the liquid, limited volume change associated to the solidification, high vapor pressure in solid state to allow sublimation under reasonable conditions of temperature and pressure, and, of course, environmental issues and price considerations. The relevant characteristics of the currently tested solvent are summarized in table 2. Among the other potential candidates, preliminary investigations [68] with polymer systems have revealed liquid carbon dioxide ($CO_2$) as a suitable solvent. The structures obtained using carbon dioxide are somewhat similar to those obtained with camphene, with a complex dendritic structure.

| Solvent | Water | Camphene | Naphthalene-Camphor | Tert-butyl alcohol |
|---|---|---|---|---|
| Solidification temperature | 0°C or lower, depending on slurry composition | 44-48°C | Naphthalene: 80°C<br>Camphor: 180°C<br>Eutectic: 31°C | 25.3°C<br>8°C for the slurry |
| Typical slurry preparation | RT | 60°C | 60°C | RT |
| temperature | | | | |
| Viscosity | 1.78 mPa.s at 0°C | 1,4 mPa.s at 47°C | Naphthalene: 0,91 mPa.s at 80°C Camphor: 0,63 mPa.s at 180°C | |
| Volume change associated to solidification | 9% | -3.1% | Negative. Depends on the composition. | 2% |
| Vapor pressure in solid state | 0,1 kPa at -20°C | 2 kPa at 55°C | Naphthalene: 0,13 kPa at 52°C Camphor: 0,13 kPa at 41°C | 6.4 kPa at 40°C |
| Usual sublimation conditions | Freeze-dryer, trap at -50°C or -85°C, pressure ? | Room temperature and atmospheric pressure | Room temperature and atmospheric pressure | 85°C, atmospheric pressure |
| Pores morphology | Lamellar channels | Dendritic channels | Dendritic channels or dense, depending on the composition | Prismatic channels |
| Environmental friendliness (Hazard Codes) | - | Highly flammable (F) | Highly flammable (F), harmfull (Xn), dangerous for the environment (N) | Highly flammable (F), harmfull (Xn) |
| Price | - | >100€/kg | >40€/kg | >300€/kg |
| Comments | Very strong anisotropy of surface tension, leading to the formation of lamellar ice crystals | | Inhibition of particles rejection with the eutectic composition (no residual porosity) | Freeze-gelcasting with acrylamide. High strength of green body. |

Table 2: Main characteristics of the solvents used for freeze-casting and resulting characteristics of the porosity.

**Ceramic powder** – The properties of the starting powder might have a major effect on the characteristics of the final materials, although little results have been reported so far. The core of the process being based on the interaction between the particles and the solidification front, a number of parameters are susceptible to modify these interactions, including the size of the particles, their size distribution, their shape, surface roughness, and surface tension. The only influence reported so far is that of the particle size [25], affecting the relationship between the solidification kinetics and the structure wavelength. Further investigations are needed to clarify these effects. The limits acceptable for the particle size are discussed in the final part.

**Slurry formulation** – As in any ceramic process, the formulation of the slurry must be carefully optimized. To ensure a homogeneous structure in the sintered materials, any segregation effects must be avoided. Slurries must be stable during the entire duration of the freezing stage. The microstructure can also be modified by varying the concentration of the starting slurry. Since the solvent initially present in the slurry is converted first into solid that is later eliminated to form the porosity, the pore content can be adjusted by tuning the slurry characteristics. The final porosity of the material is directly related to the volume of solvent in the suspension. From the data plotted in Fig. 11, it appears that a wide range of porosity content can be achieved through freeze-casting, approximately from 25 to 90%. The reasons for these limits are described in the final part of this review. The formulation has also been modified through additives, like polystyrene [26] or glycerol [12], to achieve certain effects. These additives are likely to affect the viscosity, surface tension and modify the supercooling effects. Additives can be desirable to modify the morphology of the porosity (through the shape of the solvent crystals) or modify the interaction between the particles and the solidification front. Glycerol, for example, well-known for its anti-freeze effects in other applications, was found to disrupt the morphology of the solidification front [12], and the resulting structures were dense.

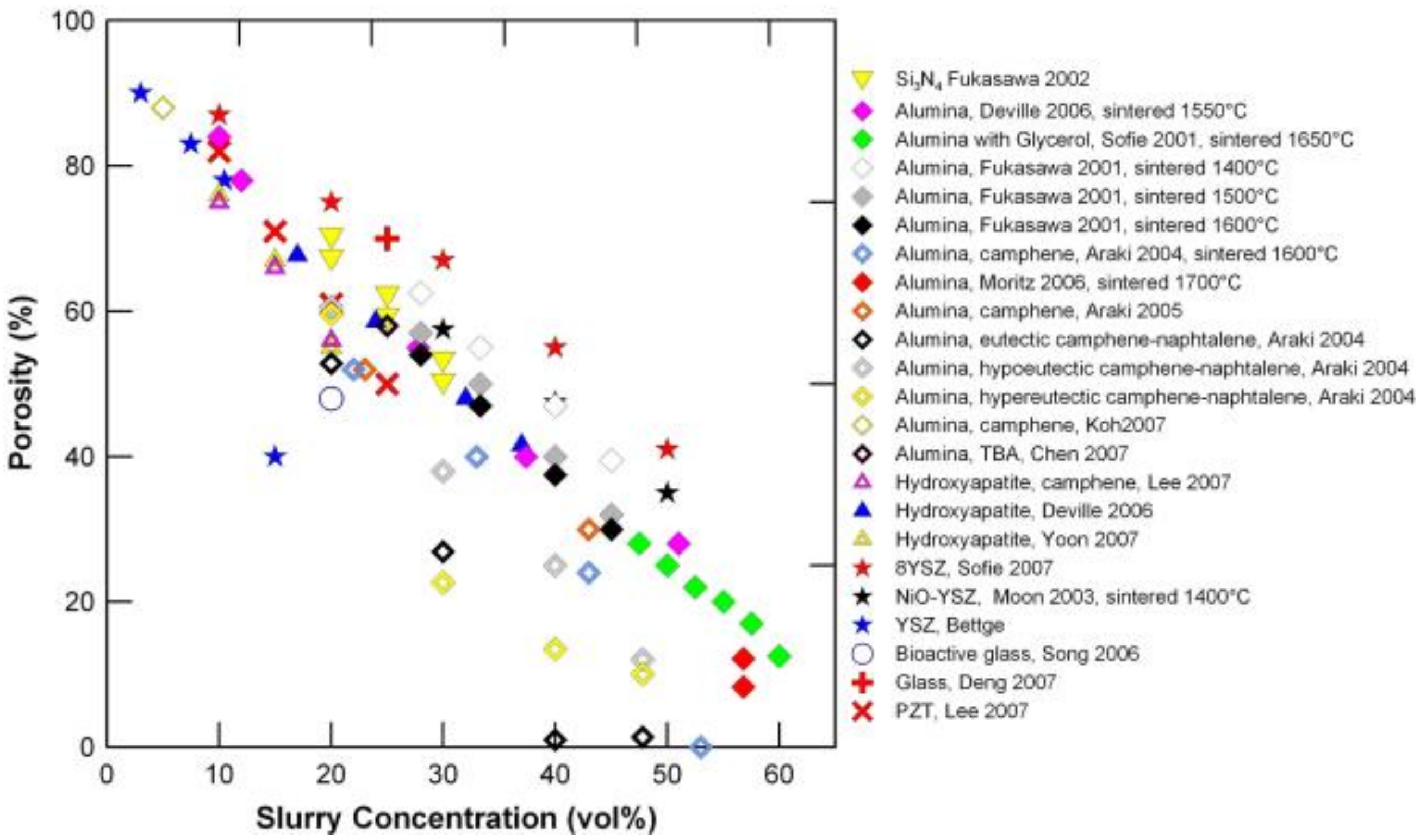

Figure 11: Total porosity vs. slurry solid loading (see Table 1 for references). No interconnected porosity was found in the studies of Sofie [12] (addition of glycerol) and Moritz [69, 70].

**Solidification conditions** – The directionality of the solidification is critical in regards of the desired directionality of the porosity. When the slurries are solidified without any temperature gradient applied, the crystals can nucleate at any place and have no preferred growth direction. This results in structures with a random orientation of the porosity. However, when the experimental setup allows imposing a defined temperature gradient, the solvent crystals are forced to grow along the temperature gradient. Provided the temperature field is carefully controlled, crystals and hence resulting pore channels can run through the entire samples, reaching dimensions of a few centimeters. The tortuosity of the porosity in such cases is close to 1. Besides the directionality, the nucleation conditions can be important. If the slurry is partially quenched, i.e., poured over a cold finger maintained at a constant and negative temperature, the initial freezing is not steady. Although lamellae and channels are observed all over the sample, their orientation over the cross-section parallel to the ice front is completely random. Colonies of locally aligned pores are observed, but no long-range order is found (see Fig. 4 of reference [25]). Homogeneous freezing (i.e., cooling of the fingers at constant rate starting from room temperature) results in a more homogeneous ice nucleation leading to an oriented and continuous lamellar porous architecture, with long range order, both in the parallel and perpendicular directions of the ice front. Solidification kinetics was also found to have a dramatic influence on the structure. When the freezing kinetics is increased, i.e., the solidification front speed increases, the width of the channels and of the lamellae is drastically affected (Fig. 12). The faster the freezing rate, the finer is the microstructure. The empirical dependence between the wavelength $\lambda$ (or the lamellae thickness) and the speed of the ice front in the direction parallel to the temperature gradient ($v$) can be described with a simple power law ($\lambda = A.v^{-n}$).

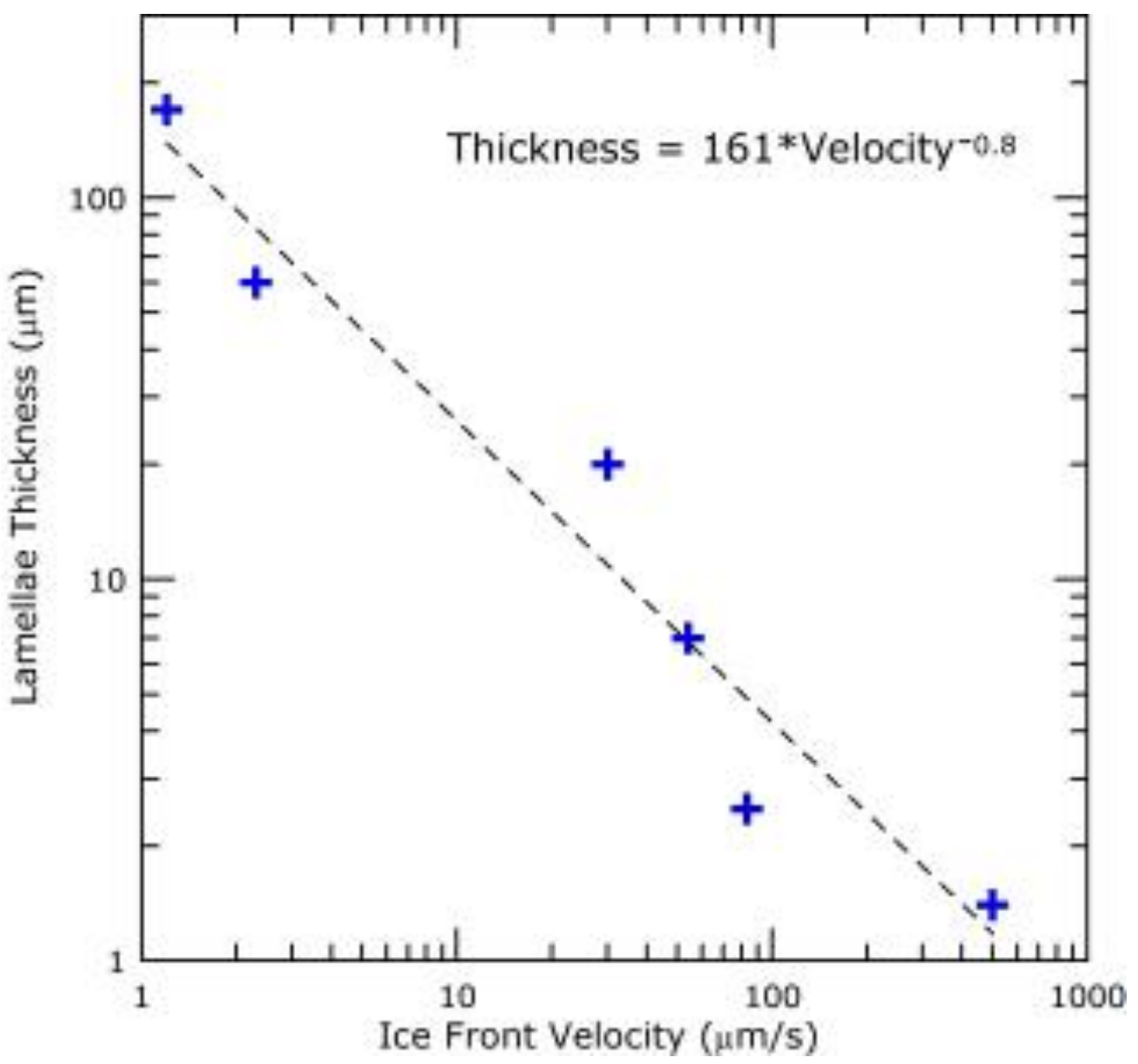

Figure 12 : Lamellae thickness vs. ice front velocity, porous lamellar alumina (after reference [17]).

**Sintering** – As with any other ceramic process, the sintering stage can be optimized to control the porosity/density of the final piece [21]. Besides the densification of the walls, the effect of sintering on the structure and macroporosity characteristics has not been tackled yet, and the mechanisms controlling the shrinkage have not been identified. The only modification reported is the formation of elongated fibrous grains during the sintering of freeze-casted porous silicon nitride, which was believed to be due to some vapor-solid phase transformation.

## 6. Limits

### *6.1 Solid content in slurry and green body density*

If the particles are rejected by the solidification front, the particle concentration in the remaining melt areas will become larger. Dendrites of the solidified solvent grow into the liquid, pushing the ceramic particles into the interdendritic spaces. Eventually, the particle redistribution ceases, and the solid/liquid interface moves into the interparticles spaces (Fig. 13). This phenomenon has been called a ‘‘breakthrough’’ [55] as the solid/liquid interface breaks into the suspension. A simple yet efficient model has been developed by Shanti et al. [55], taking into account the capillary force pushing the particles with the interface and the countering force, i.e., the resultant of the osmotic pressure of the suspension. The particle volume fraction at breakthrough was found to be determined by the maximum packing of particles at the point of jamming, modified by a small term dependent upon particle size and surface tension. The predictions agreed with observation for the alumina–camphene system within about 4%, which is quite satisfying considering the simplicity of the model used. As a consequence, the density of the green body after sublimation is always the same, providing the formulation of the slurry is kept constant. Little influence of the nature of the solvent and characteristics of the particles was found. Another consequence of this analysis is the existence of a maximum solid content in the formulation of the slurry. If the solid content already exceeds the breakthrough concentration, the solid/liquid interface moves into the interparticles space before any particle redistribution can occur. Particles distribution in the frozen samples is homogeneous, and the green body obtained after sublimation will not exhibit any continuous porosity resulting from particles-free dendrites.

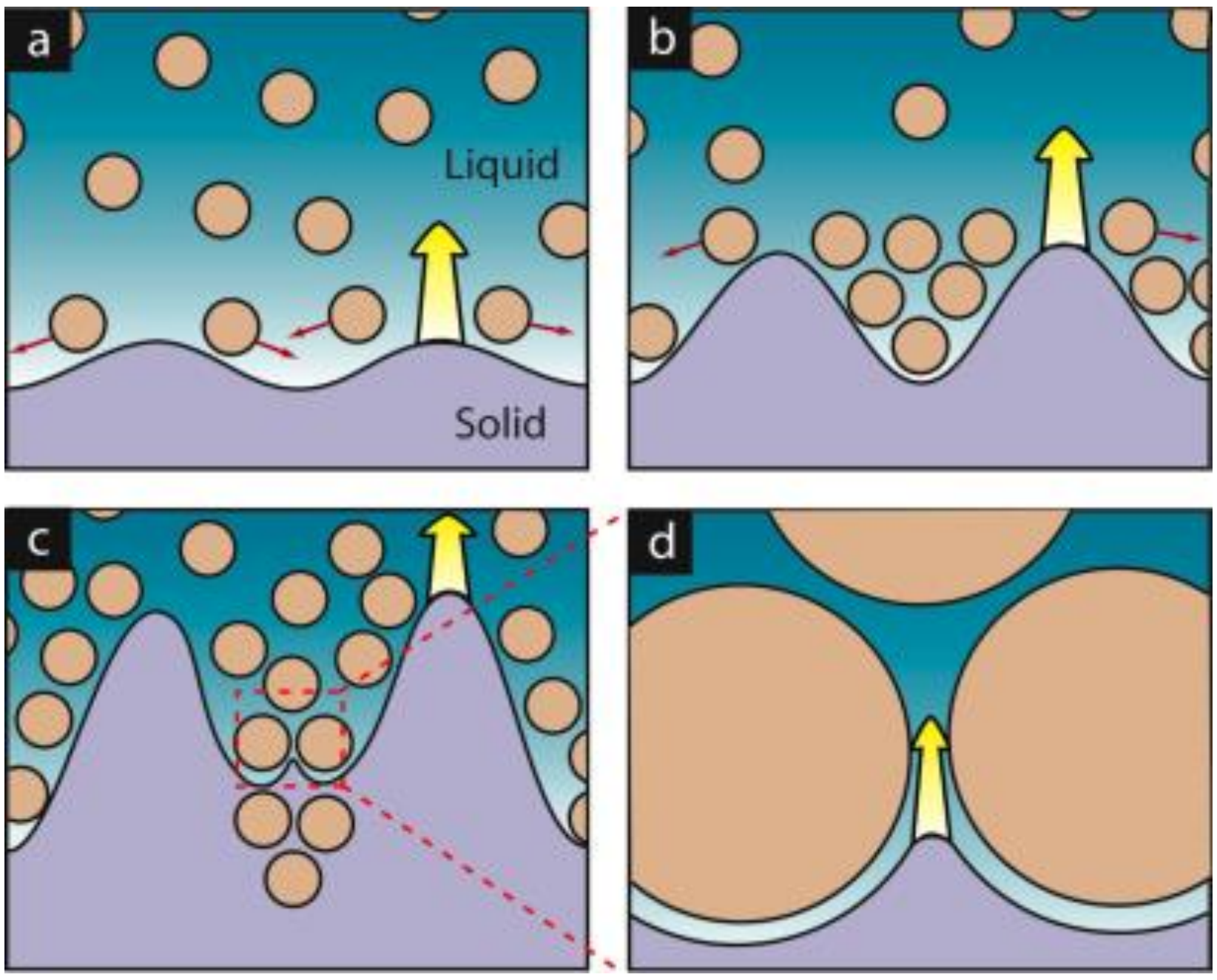

Figure 13: Particles redistribution during solidification. Particles are concentrated between the solidified dendrites (a-b) until the capillary pressure exceeds the osmotic pressure (c), at which point the solid/liquid interface moves into the interparticles space (d).

### *6.2 Directionality of the porosity*

The directionality of the porosity can be controlled by the solidification conditions, unidirectional solidification yielding similarly oriented structures. Refinements have nonetheless be developed, such as solidification in a radial direction [20], to process ceramic tubes with a radial porosity, a feature making them useful as filters (Fig. 6). Inducing a complex orientation of the porosity will directly depend on the ability to properly and homogeneously control the direction of propagation of the solidification front, becoming thus an experimental setup issue. Inspiration can also be taken from investigations of porous polymers obtained through the same approach, for example where complex interpenetrating materials were obtained. This was done by freezing in the same direction as the original aligned channels or perpendicular to it. Rather unique structures were obtained with grid-like morphologies [62]. Some of these concepts could likely be extended to porous ceramics.

### 6.3 *Walls thickness, pore size and solidification kinetics*

A number of studies [17, 23, 25, 30, 32, 47, 55] have revealed the relationships between the solidification kinetics (i.e., the solidification front speed) and the size of the porosity (or the structure) of the freeze-casted ceramics. Such relationships are commonly found in processing techniques based on directional solidification, including metallic and polymeric [62] materials. Criteria were developed to predict the dendritic spacing as a function of the solidification kinetics, and several experimental parameters have been identified. The faster the freezing rate, the finer the resulting structure, as shown before. Limits are nevertheless encountered in the case of porous ceramics, due to the underlying phenomena: the solidification of the solvent and the interaction between the solidification front and the ceramic particles. Preliminary results can be found in reference [25], although a lot of work is still necessary to investigate these limits. One limit is related to the critical velocity for entrapment [71]. Above this critical velocity, particles are entrapped by the moving interface and porosity is disappearing from the final structure. Upgrading the experimental setup to increase the freezing kinetics will be helpless with this issue beyond a certain point, and obtaining cooling rate in the range of these obtained during the solidification of metallic alloys will be useless. Several parameters seem nonetheless to affect this critical velocity, the most important being probably the solid content in the slurry, the particle size and the properties of the solvent. These limits are schematically summarized in Fig. 14.

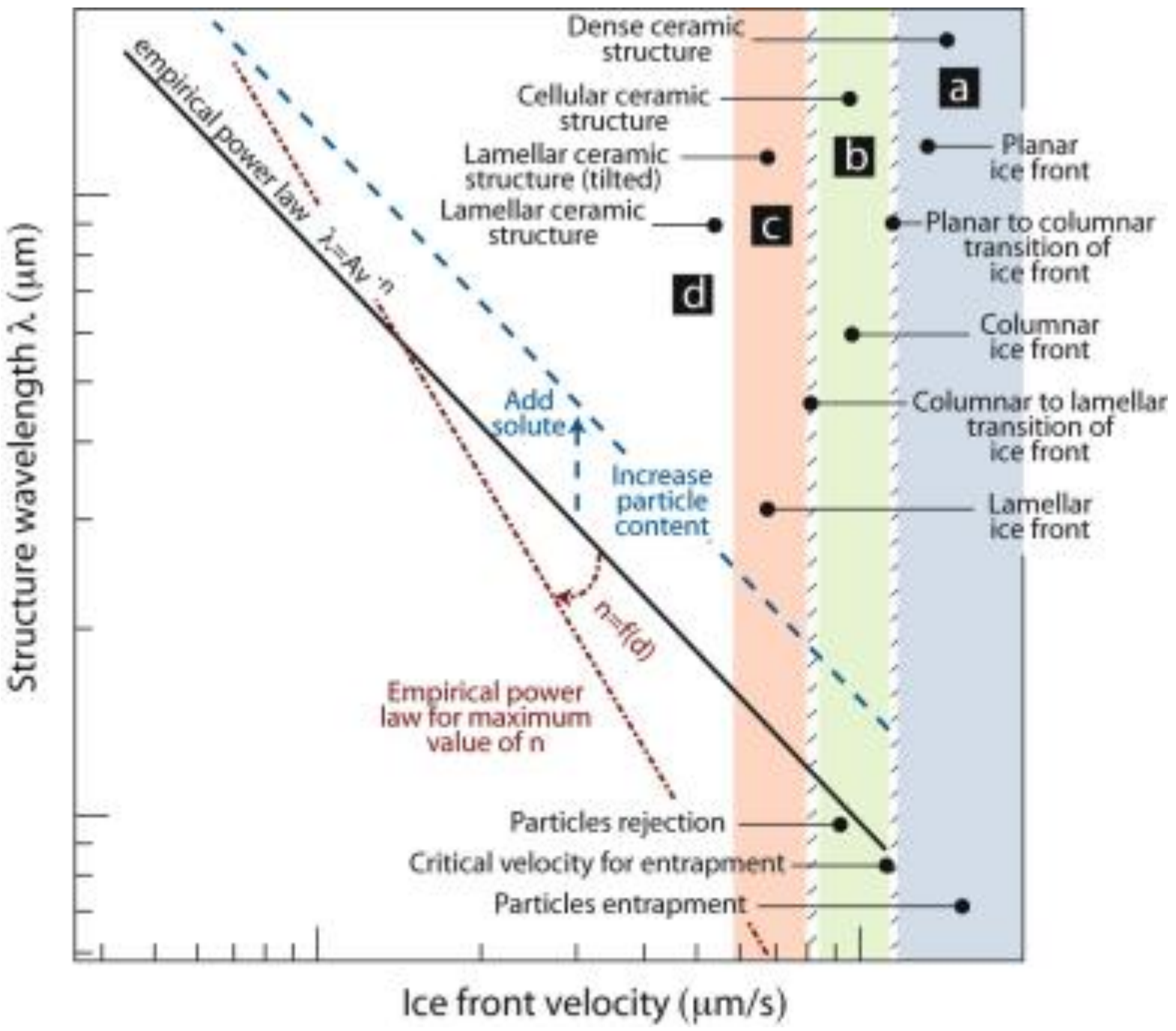

Figure 14 : Strategies and limits for controlling the structure: schematic plot of wavelength vs. ice front velocity [25]. The exponent n of the empirical law is dependent on the particle size *d*, though the function $n = f(d)$ is not monotonic; an optimum of $d$ is encountered where the exponent $n$ is maximum. Very fast cooling rates, region (a), will result in the ice front trapping the particles and the formation of a dense material. When the velocity is decreased below the critical value for particle entrapment, $v_c$, the particles are expelled from the growing ice but if the speeds are fast enough the ice will grow with a columnar microstructure, region (b). Slower velocities will result in the formation of lamellar ice. However, if the velocity is still fast or, equivalently, the gradient in temperature small enough, the balance between the preferential growth direction and the gradient direction will result in the growth of lamellae tilted with respect to the later, region (c). As the velocity decreases (or the gradient increases) the lamellae will align with the direction of the temperature gradient, region (d).

### *6.4 Pores morphology*

The morphology and characteristics of the pores are influenced by several independent or interconnected parameters, as shown previously. Any modification of these parameters will have a direct repercussion on the structure of the materials. Three main parameters have nevertheless been identified as being critical with regards to this issue:

**Nature of the solvent** – This one is probably the most critical one. The crystallographic properties of the solvent in its solid state will define the main appearance of the structure, i.e., lamellar with water, dendritic with camphene or prismatic with tert-butyl alcohol [54]. Obtaining radically different morphologies will imply a new choice of solvent with different properties, and will be restricted by the requirements exposed previously.

**Freezing conditions** – Both the kinetics and the directionality of the freezing conditions can be used to modify the pores morphology. For the reasons previously exposed, the freezing conditions will affect both the pore size and the pore morphology, so that these two parameters are usually interdependent.

**Particle size** – The homogeneity of the porous structure is lost when the particle size become too similar to the size of the solvent crystals. Morphological features of the crystals cannot be well replicated into the final structure if the size of the particle is in the range of order of that of the interdendritic spaces. Hence, the particle size must be kept below the wavelength of the desired structure. Micron-sized particles can obviously not be used to create nanometer-sized structure (Fig. 15). Freeze-casting seem to be appealing for creating porous structures with a pore size smaller than a hundred microns. In that case, to ensure a homogeneous and well defined structure, powders with submicronic mean particle size seem to be desirable. Distribution of the particle size should also be taken into account for the same reasons, and the presence of large particle or particles agglomerates is detrimental to the homogeneity of the final structure. Finally, the critical velocity for particles entrapment is inversely proportional to the particle size. Using larger particles will lower this critical velocity and processing conditions might fall in the range were fast freezing kinetics (desirable to get small pores) cannot be used.

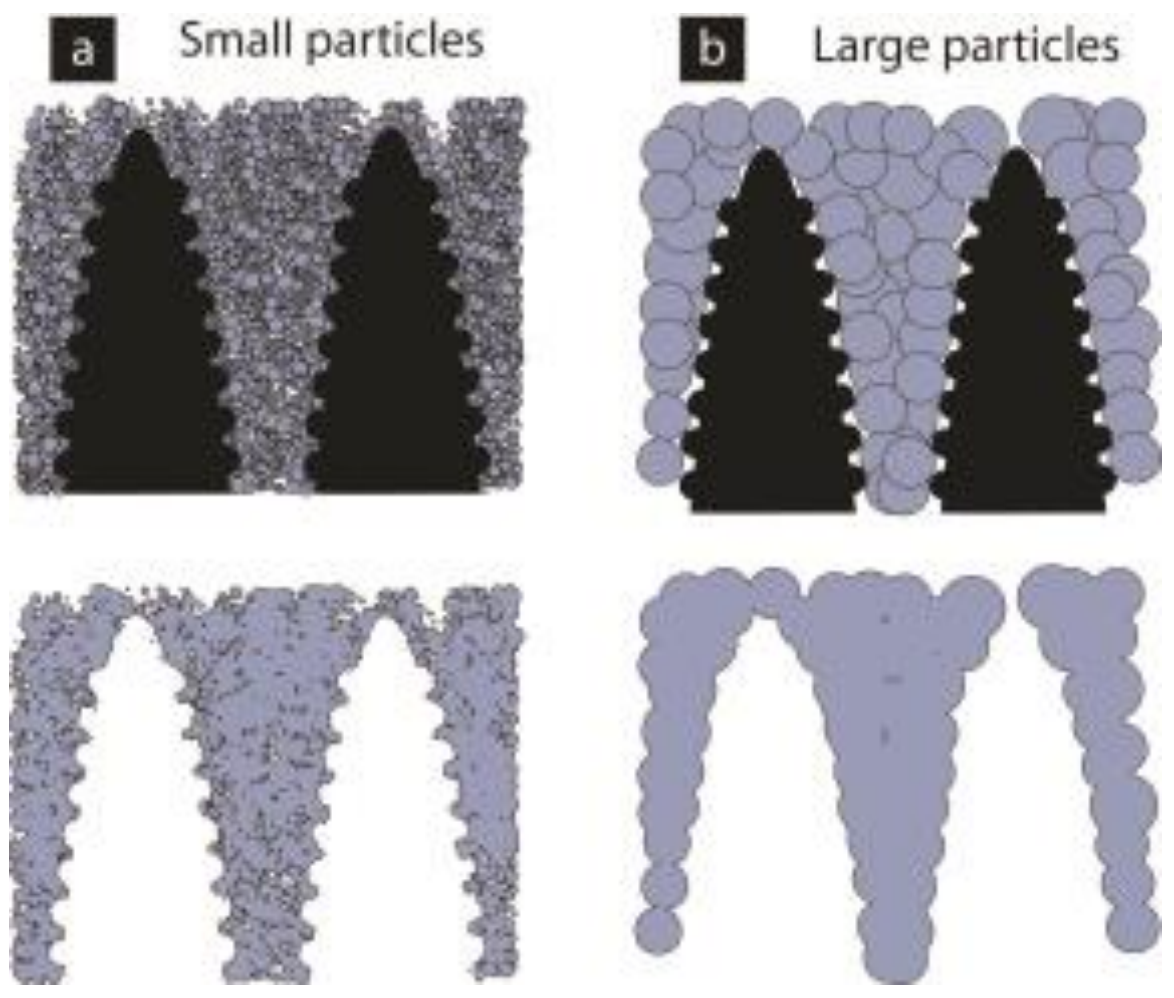

Figure 15: Influence of particle size on pores morphology. Details of the crystals are well replicated in the case of small particles (a) whereas such details are lost with large particles (b).

### *6.5 Processing time and dimensions of the pieces*

Two stages of the process are critical in regards of the total processing time and the dimensions of the final pieces: the freezing stage and the sublimation stage.

**Freezing time** – The freezing time required will depend on two parameters: the desired structure wavelength (i.e., pore size or wall thickness) and the dimensions of the final piece. Regarding the structure wavelength, freeze-casting seem to offer unique opportunities for dimensions below a hundred microns. Within this range, the freezing kinetics required are rapid enough so that the processing times remain reasonable (a few minutes for pieces of a few centimeters). Reaching smaller pores will imply faster freezing, so that the freezing time is even further reduced. The desired dimensions of the pieces must also be taken into account. Most of the investigations reported so far were dealing with sample's thickness in the range 1-3 cm, i.e., at the lab scale. Freezing time will more or less linearly increase with the thickness of the sample, but maintaining the homogeneity of the temperature field and the freezing kinetics (which control the homogeneity of the porosity) will become problematic for pieces of larger dimensions. This will imply both a better control of the formulation to ensure its stability during the solidification and improved experimental setup with optimal temperature control.

**Sublimation** – The time necessary for sublimating the solvent is directly dependent on the dimensions of the pieces to be treated. While kept in the centimeter range, sublimation can occur overnight. This stage must nevertheless be carefully controlled to avoid any formation of defects (cracks, distortion) during the solid-vapor phase transition, and such control might become problematic for larger pieces. Technical solutions are already available to treat pieces of larger dimensions.

## 7. Applications of porous freeze-casted ceramics

The potentialities offered by the unique structures obtained by freeze-casting have drawn attention for a number of very diverse applications, although very few figures of functional properties have been reported so far. The applications considered so far belongs to two main groups:

**Biomaterials** – [17, 24, 30-35, 45, 69, 70]. The structure of the porous ceramics obtained by freeze-casting being very similar to that of natural biomaterials such as nacre or bone [72], biomaterials has become a field with potentially important applications of this technology, with particular attention to tissue engineering. Preliminary reports of dramatic improvements of the compressive strength of freeze-casted hydroxyapatite seem to confirm that potential [17], and the assessment of the biological response of these materials is under way in various laboratories.

**Materials for chemical processes and energy sources** – including SOFC, electrodes, catalysts, sensors, filtration/separation devices [13, 16, 18-21, 37, 38, 40, 41, 43, 47, 48, 52, 56, 73, 74], and photocatalysis for air or water purification [44]. The unique morphology of the porosity is of interest for these applications, where open structures with good connectivity and low tortuosity (close to 1, ensuring optimal access of the fluids to the reaction sites), large surface area (large number of electrochemical reaction sites), dense ceramic walls (good conductivity) and good mechanical properties are required. The directionality of the porosity can be used to optimize the species transport properties while the dense walls will ensure optimal conducting properties and mechanical reliability. The versatility of the process allows considering a wide range of technical ceramics or composites (e.g., NiO-YSZ), and the typical dimensions of the samples processed so far appear to be suitable for most of these applications.

A recent report of freeze-casted piezoelectrics [42] can also be found, with high hydrostatic figures of merit. Such materials could find applications as actuators, transducers and in particular low frequency hydrophones.

## 8. Perspectives

Following a promising beginning, a bright future seems to open for freeze-casted porous ceramics. Perspectives for the technology can be classified along two axes: the development of the technique and its underlying science. The present review has highlighted strategies and limits for improving the structure of the materials and the experimental setups. Incremental developments can be expected in the following aspects (although things never turn out quite the way you expect, of course…):

- Improved control of the structure by improvements of the setup, in particular with regards to the homogeneity of the temperature field and the directionality of freezing. Inspiration can be taken from setups developed for freeze-casting of polymers, such as the power-down setup used to process porous collagen plates [15]. The improvement of the setup should probably consider the issues associated with the scaling up of the process.
- Development of functional structures, either by an improvement of the experimental setup, such as the recent freeze-tape casting developments [16, 39], or by a post-processing functionalisation of the structures, as recently reported with the nano-$TiO_2$ coating [44] deposited on freeze-casted porous ceramics, to add a photocatalysis function. Such functionalisation will be required for the majority of the applications considered so far, from biomaterials scaffolds to SOFC.

- The development of functional structure will be probably benefit from the development of new materials and corresponding processing strategies, such as the silicon carbide obtained after pyrolysis of freeze-casted precursors [52], or complex composites (polymer/ceramic, metal/ceramic, ceramic/ceramic), obtained either in one-step [62] or two-steps approaches [17].

From a fundamental point of view, a large number of aspects of the processes are still to be fully understood. The core problem of the process, the particles/solidification front interactions, is complex to understand and model, and both theoretical and experimental work will be welcome. Preliminary modeling work has been undertaken [75] and should be highly beneficial to control and optimizes the structure of freeze-casted porous ceramics. This analysis could largely benefit from the knowledge derived from other fields involved with the same underlying science.

**References**

[1] D. J. Green, R. Colombo, *MRS Bull.* **2003**, *28*, 296.
[2] L. J. Gibson, *MRS Bull.* **2003**, *28*, 270.
[3] A. Herzog, R. Klingner, U. Vogt, T. Graule, *J. Am. Ceram. Soc.* **2004**, *87*, 784.
[4] H. Sieber, C. Hoffmann, A. Kaindl, P. Greil, *Adv. Eng. Mater.* **2000**, *2*, 105.
[5] B. Sun, T. Fan, D. Zhang, *J. Porous Mater.* **2002**, *9*, 275.
[6] J. Cao, C. R. Rambo, H. Sieber, *J. Porous Mater.* **2004**, *11*, 163.
[7] P. Sepulveda, J. G. P. Binner, *J. Eur. Ceram. Soc.* **1999**, *19*, 2059.
[8] O. O. Omatete, M. Janney, *Am. Ceram. Soc. Bull.* **1991**, *70*, 1641.
[9] G. Tari, *Am. Ceram. Soc. Bull.* **2003**, *82*, 43.
[10] U. T. Gonzenbach, A. R. Studart, E. Tervoort, L. J. Gauckler, *J. Am. Ceram. Soc.* **2007**, *90*, 16.
[11] H. Lu, Z. Qu, Y. C. Zhou, *J. Mater. Sci.-Mater. M* **1998**, *9*, 583.
[12] S. W. Sofie, F. Dogan, *J. Am. Ceram. Soc.* **2001**, *84*, 1459.
[13] T. Fukasawa, M. Ando, T. Ohji, S. Kanzaki, *J. Am. Ceram. Soc.* **2001**, *84*, 230.
[14] J. S. Wettlaufer, M. G. Worster, H. E. Huppert, *J. Fluid Mech.* **1997**, *344*, 291.
[15] H. Schoof, J. Apel, I. Heschel, G. Rau, *J Biomed Mater Res* **2001**, *58*, 352.
[16] S. W. Sofie, *J. Am. Ceram. Soc.* **2007**, *90*, 2024.
[17] S. Deville, E. Saiz, R. K. Nalla, A. Tomsia, *Science* **2006**, *311*, 515.
[18] S. R. Mukai, H. Nishihara, H. Tamon, *Chem. Commun.* **2004**, 874.
[19] J.-W. Moon, H. J. Hwang, M. Awano, K. Maeda, S. Kanzaki, *J. Ceram. Soc. Jpn.* **2002**, *10*, 479.
[20] J.-W. Moon, H.-J. Hwang, M. Awano, K. Maeda, *Mater. Lett.* **2003**, *57*, 1428.
[21] T. Fukasawa, Z. Y. Deng, M. Ando, T. Ohji, Y. Goto, *J. Mater. Sci.* **2001**, *36*, 2523.
[22] D. Donchev, N. Walther, J. Ulrich, *VDI Berichte* **2005**, 337.
[23] D. Koch, L. Andresen, T. Schmedders, G. Grathwohl, *J. Sol-Gel Sci. Technol.* **2003**, *26*, 149.
[24] K. Araki, J. W. Halloran, *J. Am. Ceram. Soc.* **2005**, *88*, 1108.
[25] S. Deville, E. Saiz, A. P. Tomsia, *Acta Mater.* **2007**, *55*, 1965.
[26] Y.-H. Koh, E.-J. Lee, B.-H. Yoon, J.-H. Song, H.-E. Kim, H.-W. Kim, *J. Am. Ceram. Soc.* **2006**, *89*, 3646.
[27] M. Nakata, K. Tanihata, S. Yamaguchi, K. Suganuma, *J. Ceram. Soc. Jpn.* **2005**, *113*, 712.
[28] K. Araki, J. W. Halloran, *J. Am. Ceram. Soc.* **2004**, *87*, 2014.
[29] B.-H. Yoon, C.-S. Park, H.-E. Kim, Y.-H. Koh, *Mater. Lett.*, *In Press, Corrected Proof.*

[30] S. Deville, E. Saiz, A. Tomsia, *Biomaterials* **2006**, *27*, 5480.
[31] E.-J. Lee, Y.-H. Koh, B.-H. Yoon, H.-E. Kim, H.-W. Kim, *Mater. Lett.* **2007**, *61*, 2270.
[32] B.-H. Yoon, Y.-H. Koh, C.-S. Park, H.-E. Kim, *J. Am. Ceram. Soc.* **2007**, *90*, 1744.
[33] T. Moritz, H.-J. Richter, *J. Eur. Ceram. Soc.* **2007**, *27*, 4595.
[34] S. Deville, P. Miranda, E. Saiz, A. P. Tomsia, "Fabrication of porous hydroxyapatite scaffolds", presented at *International Conference on Manufacturing Science and Engineering*, Ypsilanti, MI, **2006**.
[35] S. Deville, E. Saiz, R. K. Nalla, A. P. Tomsia, "Strong Biomimetic Hydroxyapatite Scaffolds", presented at *4th Forum on New Materials*, Acireale, Sicily, **2006**.
[36] Y. Suetsugu, Y. Hotta, M. Iwasashi, M. Sakane, M. Kikuchi, T. Ikoma, T. Higaki, N. Ochiai, J. Tanaka, *Key Eng. Mater.* **2007**, *330-332 II*, 1003.
[37] Y. H. Koh, J. J. Sun, H. E. Kim, *Mater. Lett.* **2007**, *61*, 1283.
[38] M. Bettge, H. Niculescu, P. J. Gielisse, "Engineered porous ceramics using a directional freeze-drying process", presented at *28th International Spring Seminar on Electronics Technology*, **2005**.
[39] L. Ren, Y.-P. Zeng, D. Jiang, *J. Am. Ceram. Soc.* **2007**, *90*, 3001.
[40] T. Fukasawa, M. Ando, T. Ohji, *J. Ceram. Soc. Jpn.* **2002**, *110*, 627.
[41] T. Fukasawa, Z. Y. Deng, M. Ando, T. Ohji, S. Kanzaki, *J. Am. Ceram. Soc.* **2002**, *85*, 2151.
[42] S.-H. Lee, S.-H. Jun, H.-E. Kim, Y.-H. Koh, *J. Am. Ceram. Soc.* **2007**, *90*, 2807.
[43] S. Ding, Y.-P. Zeng, D. Jiang, *J. Am. Ceram. Soc.* **2007**, *90*, 2276.
[44] Z.-Y. Deng, H. R. Fernandes, J. M. Ventura, S. Kannan, J. M. F. Ferreira, *J. Am. Ceram. Soc.* **2007**, in press.
[45] J.-H. Song, Y.-H. Koh, H.-E. Kim, L.-H. Li, H.-J. Bahn, *J. Am. Ceram. Soc.* **2006**, *89*, 2649.
[46] P. Kisa, P. Fisher, A. Olszewski, I. Nettleship, N. G. Eror, "Synthesis of porous ceramics through directional solidification and freeze-drying", presented at *Materials Research Society Symposium*, Boston, MA., **2003**.
[47] H. Nishihara, S. R. Mukai, D. Yamashita, H. Tamon, *Chem. Mater.* **2005**, *17*, 683.
[48] H. Nishihara, S. R. Mukai, Y. Fujii, T. Tago, T. Masuda, H. Tamon, *J. Mater. Chem.* **2006**, *16*, 3231.
[49] H. J. Hwang, D. Y. Kim, J.-W. Moon, *Key Eng. Mater.* **2006**, *510-511*, 906.
[50] J. Tang, Y. Chen, H. Wang, H. Liu, Q. Fan, *Key Eng. Mater.* **2005**, *280-283*, 1287.
[51] S. Yunoki, T. Ikoma, A. Monkawa, K. Ohta, M. Kikuchi, S. Sotome, K. Shinomiya, J. Tanaka, *Mater. Lett.* **2006**, *60*, 999.
[52] B.-H. Yoon, E.-J. Lee, H.-E. Kim, Y.-H. Koh, *J. Am. Ceram. Soc.* **2007**, *90*, 1753.
[53] B.-H. Yoon, C.-S. Park, H.-E. Kim, Y.-H. Koh, *J. Am. Ceram. Soc.*, *0*, ???
[54] R. Chen, C.-A. Wang, Y. Huang, L. Ma, W. Lin, *J. Am. Ceram. Soc., In Press, Corrected Proof.*
[55] N. O. Shanti, K. Araki, J. W. Halloran, *J. Am. Ceram. Soc.* **2006**, *89*, 2444.
[56] Y. H. Koh, I. K. Jun, J. J. Sun, H. E. Kim, *J. Am. Ceram. Soc.* **2006**, *89*, 763.
[57] J. A. Sekhar, R. Trivedi, *Mater. Sci. Eng., A* **1991**, *A147*, 9.
[58] J. D. Hunt, *Mater Sci Tech-Lond* **1999**, *15*, 9.
[59] J. O. M. Karlsson, *Science* **2002**, *296*, 655.
[60] G. Gay, M. A. Azouni, *Cryst. Growth Des.* **2002**, *2*, 135.
[61] D. K. Perovich, A. J. Gow, *J. Geophys. Res. C* **1996**, *101*, 18327.
[62] H. F. Zhang, I. Hussain, M. Brust, M. F. Butler, S. P. Rannard, A. I. Cooper, *Nature Materials* **2005**, *4*, 787.
[63] C. Korber, Rau, G., Cosman, M.D., Cravalho, E.G., *J. Cryst. Growth* **1985**, *72*, 649.
[64] H. Ishiguro, B. Rubinsky, *Cryobiology* **1994**, *31*, 483.
[65] W. W. Mullins, R. F. Sekerka, *J. Appl. Phys.* **1964**, *35*, 444.
[66] L. Hadji, *Eur Phys J B* **2004**, *37*, 85.

[67] L. Hadji, A. M. J. Davis, *J. Cryst. Growth* **1998**, *191*, 889.
[68] H. Zhang, J. Long, A. I. Cooper, *J. Am. Chem. Soc.* **2005**, *127*, 13482.
[69] T. Moritz, H.-J. Richter, *J. Am. Ceram. Soc.* **2006**, *89*, 2394.
[70] T. Moritz, H.-J. Richter, *Adv Sci Tech* **2006**, *45*, 391.
[71] A. W. Rempel, M. G. Worster, *J. Cryst. Growth* **1999**, *205*, 427.
[72] S. Deville, E. Saiz, A. P. Tomsia, in *Handbook of Biomineralization*, Vol. 2 (Eds: P. Behrens, E. Bauerlein), Wiley-VCH, Weinheim **2007**.
[73] S. R. Mukai, H. Nishihara, H. Tamon, *Microporous Mesoporous Mater.* **2003**, *63*, 43.
[74] T. Fukasawa, Z. Y. Deng, M. Ando, T. Ohji, "Porous ceramics with controlled aligned-pores synthesized by freeze-drying process", presented at *25th Annual Conference on Composites*, Cocoa Beach, FL, **2001**.
[75] S. S. L. Peppin, J. A. W. Elliot, M. G. Worster, *J. Fluid Mech.* **2006**, *554*, 147.